\documentclass[12pt]{iopart}
\usepackage{amssymb, amsthm,textcomp,enumerate}
\usepackage{graphicx,color}
\usepackage{verbatim}
\usepackage{xspace}
\usepackage{amsfonts}
\usepackage{cancel}

\newcommand{\hmol}{$\mathrm{H}_{2}$\xspace}
\newcommand{\ehmol}{$\left(e^{+}-\mathrm{H}_{2}\right)$\xspace}
\newcommand{\deter}{$\mathrm{det}\left(A\right)$\xspace}
\newcommand{\deterc}{$\mathrm{det}\left(A'\right)$\xspace}
\newcommand{\targ}{$\psi_{\mathrm{G}}$\xspace}
\newcommand{\ps}{$\eta_{\mathrm{v}}$\xspace}
\newcommand{\psc}{$\eta'_{\mathrm{v}}$\xspace}
\newcommand{\pst}{$\eta_{\mathrm{v}}\left(\tau\right)$\xspace}
\newcommand{\trialwave}{$\Psi_{\mathrm{t}}$\xspace}
\newcommand{\ctrialwave}{$\breve{\Psi}_{\mathrm{t}}$\xspace}
\newcommand{\fderiv}{$\partial\eta_{\mathrm{v}}/\partial\tau$\xspace}
\newcommand{\cderiv}{$\partial\eta'_{\mathrm{v}}/\partial\tau$\xspace}
\newcommand{\kderiv}{$\partial\Re\left[\eta'_{\mathrm{v}}\right]/\partial k$\xspace}
\newcommand{\sderiv}{$\partial^{2}\eta_{\mathrm{v}}/\partial\tau^{2}$\xspace}
\newcommand{\tderiv}{$\partial^{3}\eta_{\mathrm{v}}/\partial\tau^{3}$\xspace}

\begin{document}
 
\title[Equivalence of the generalized and complex Kohn variational methods]{Equivalence of the generalized and complex Kohn variational methods}

\author{J N Cooper$^1$, M Plummer$^2$ and E A G Armour$^1$}
\address{$^1$ School of Mathematical Sciences, University Park, Nottingham NG7 2RD, UK}
\address{$^2$ STFC Daresbury Laboratory, Daresbury, Warrington, Cheshire WA4 4AD, UK}
\ead{pmxjnc@googlemail.com}

\begin{abstract}
For Kohn variational calculations on low energy \ehmol elastic scattering, we
prove that the phase shift approximation obtained using the complex Kohn
method is precisely equal to a value which can be obtained immediately via the
real-generalized Kohn method. Our treatment is sufficiently
general to be applied directly to arbitrary potential scattering or single open channel scattering problems, with exchange if
required. In the course of our analysis, we develop a
framework formally to describe the anomalous behaviour of our generalized Kohn
calculations in the regions of the well known Schwartz singularities. This
framework also explains the mathematical origin of the anomaly-free
singularities we reported in a previous article. Moreover, we demonstrate a novelty, that
explicit solutions of the Kohn equations are not required in order to
calculate optimal phase shift approximations. We relate our rigorous framework to earlier descriptions
of the Kohn-type methods.
\end{abstract}

\pacs{02.10.Yn, 34.10.+x, 34.80.Bm, 34.80.Uv}
\submitto{J. Phys. A: Math. Theor.}
\maketitle

\section{Introduction}\label{s:introduction}

In a recent article \cite{CooperArmourPlummer2009}, we presented variational
calculations of phase shifts for the elastic scattering of very low energy
positrons, $e^{+}$, by ground state molecular hydrogen, \hmol. Those
calculations involved modifications of the variational method originally due
to Kohn \cite{Kohn1948} and Hulth\'en \cite{Hulthen1944,Hulthen1948} (see also
Rubinow \cite{Rubinow1955}), which is the analogue of the Rayleigh-Ritz method
\cite{BransdenJoachain2003} widely used in variational calculations on bound
states. Our earlier analysis concentrated on the so-called Schwartz
singularities \cite{Schwartz1961,Schwartz1961b}, which are well known to be
associated with anomalous behaviour in the results of Kohn
calculations. In our investigation of \ehmol scattering, we considered the
generalization of the Kohn method due to Kato \cite{Kato1950,Kato1951} as well
as the complex Kohn method \cite{McCurdy1987,Schneider1988}. Both of these
modifications are designed to circumvent problems associated with
Schwartz-type anomalous behaviour; the latter especially has become popular in
recent years as it has been commonly believed automatically to be free of all 
nonphysical anomalies. 

We have carried out Kohn calculations for \ehmol scattering in the fixed-nuclei
approximation \cite{Temkin1967,Temkin1969} so that only the motion of the
three light particles need be considered. All of our calculations use
Hartree atomic units. Moreover, we regard the positron as
being of sufficiently low energy that elastic scattering is the only
significant open channel other than positron annihilation and that only the
lowest partial wave of $\Sigma_{\mathrm{g}}^{+}$ symmetry need be
investigated. Armour and coworkers \cite{Armour1990} have shown that
consideration of this partial wave alone is sufficient to account for \ehmol
scattering processes up to $\sim2$ eV. This partial wave is equivalent to the s-wave in atomic scattering.

In this article, we will construct a framework ultimately to describe a formal connection between the mechanics of the real-generalized and complex Kohn methods. We will prove that the phase shift approximation determined in the complex Kohn method is exactly equal to that obtained from the real-generalized Kohn method (hereafter referred to as the generalized Kohn method, for brevity) under a particular optimization.
In developing the framework leading to this conclusion, we will establish
three other important results. Firstly, we will identify the mathematical
origin of a certain class of anomalies as they appear in generalized Kohn
calculations, before proving that such anomalies are necessarily absent in
applications of the complex Kohn method. Secondly, we will show that the
existence of the anomaly-free singularities we reported earlier
\cite{CooperArmourPlummer2009} is fully accounted for by this
framework. Finally, for both the generalized and complex Kohn methods, we will demonstrate that explicit solutions of the variational equations are not required to obtain optimized phase shift approximations, since these
can be determined generally from the evaluation of four
matrix determinants. The optimized results show that the generalized and complex 
Kohn methods are equivalent.

The treatment given here for \ehmol scattering is sufficiently general that it
should not be difficult to adapt it to other physical systems. Indeed, it
applies directly to potential scattering problems and single open channel
scattering problems, with exchange if required. We will, therefore,
concentrate on abstract mechanisms rather than presenting numerical data from
individual calculations. Following the analysis, we will briefly relate our
framework to earlier work on avoiding the Schwartz singularities and the more 
familiar notation generally used by other authors (see, for example, Burke \cite{Burke1977}, and Burke and Joachain
\cite{BurkeJoachain1995}, with multichannel extensions summarized by Nesbet
\cite{Nesbet1980}). We will also demonstrate why we do not use this more
familiar notation throughout.

\section{Anomalous behaviour in the generalized Kohn method}\label{s:genm}
\subsection{Variational approximations to the phase shift}\label{sec:realintro}

Concerning the present calculations, our implementation of the generalized Kohn method is analogous to that described in a previous article \cite{CooperArmourPlummer2009}, so only a brief review of the basic principles will be given here. We approximate the exact leptonic scattering wavefunction, $\Psi$, with a trial wavefunction, viz.

\begin{equation}\label{eq:INT:trialwaveEXPL}
\Psi_{\mathrm{t}} = \left(\bar{S} + a_{\mathrm{t}}\bar{C}\right)\psi_{\mathrm{G}} + \sum_{i=1}^{M} p_{i}\chi_{i},
\end{equation}
where

\begin{equation}\label{eq:tautrans}
\left[ \begin{array}{c}
{\bar{S}}\\
{\bar{C}}
\end{array}\right] =
\left[ \begin{array}{cc}
\cos(\tau) & \sin(\tau)\\
-\sin(\tau) & \cos(\tau)\\
\end{array}\right]
\left[ \begin{array}{c}
{S}\\
{C}\\
\end{array}\right].
\end{equation}
As we have explained elsewhere \cite{CooperArmourPlummer2009},
$\tau\in\left[0,\pi\right)$ is a phase parameter that can be adjusted to avoid
nonphysical, anomalous behaviour in the results of Kohn calculations. The
functions, $S$ and $C$, form a basis that represents very low energy incident
and scattered positrons asymptotically far from the target molecule; they
account only for the lowest partial wave of $\Sigma_{\mathrm{g}}^{+}$ symmetry
and they are independent of $\tau$. For convenience, their explicit forms will
here be taken to be the same as those defined by equations (3) and (4) of our earlier article
\cite{CooperArmourPlummer2009}. In those calculations, prolate spheroidal coordinates were
used for their convenience in describing the two-centre \hmol system, though the general features of the following
analysis are not particular to this choice. That is to say, without any great modifications
an equivalent treatment could be given for different bases and coordinate systems better suited to other 
problems. For example, calculations on short-range radial-potential partial wave scattering
\cite{Burke1977,BurkeJoachain1995} use spherical coordinates with open channel radial
functions which are asymptotically proportional to spherical Bessel functions.

The function, \targ, is a real-valued, unit-normalized approximation to the
electronic ground state wavefunction of the unperturbed hydrogen molecule. It is evaluated at a fixed internuclear separation, typically the equilibrium value. Other than the
requirements that it is square-integrable and independent of both $\tau$ and
the coordinates of the positron, the precise form of \targ is not important
here (although an accurate description of the target is, of course, important when
calculating physically relevant results \cite{Cooper2008}). The functions, 
$\{\chi_{i}\}$, are real-valued short-range correlation
functions. They are used to describe interactions between the positron and the
target electrons when they are close together. Apart from the restrictions
that they must be independent of $\tau$ and square-integrable, thus
contributing negligibly to \trialwave when the positron is asymptotically far
from the target molecule, the precise forms of these functions are unimportant here. It should be noted that
(\ref{eq:INT:trialwaveEXPL}) does not explicitly contain the function,
$\chi_{0}$, which appeared in the Kohn trial function used earlier
\cite{CooperArmourPlummer2009}. This omission does not affect the generality
of our calculations and has been made purely to simplify the notation involved in the analysis that follows.

As before \cite{CooperArmourPlummer2009}, for brevity we will abbreviate by
$\langle X,Y\rangle$ any integral of the form $\langle
X\vert\left(\hat{H}-E\right)\vert Y\rangle$, where $\hat{H}$ is the
nonrelativistic Hamiltonian for the scattering system in the adiabatic nuclei approximation and $E$ is the sum of
the kinetic energy of the positron and the ground state energy expectation
value of \targ. Integrals of this form are evaluated over the configuration
space of the positron and the two electrons. The operator,
$\left(\hat{H}-E\right)$, contains a term of the form
$-\frac{1}{2}\left(\nabla^2+k^2\right)$, where $-\frac{1}{2}\nabla^2$ and
$k>0$ are, respectively, the kinetic energy operator and the wavenumber of the
positron. Consequently, for $\bar{S}$ and $\bar{C}$ as in
\cite{CooperArmourPlummer2009}, it is well known that

\begin{equation}\label{eq:INT:sccs}
\langle\bar{S}\psi_{\mathrm{G}},\bar{C}\psi_{\mathrm{G}}\rangle-\langle\bar{C}\psi_{\mathrm{G}},\bar{S}\psi_{\mathrm{G}}\rangle=\tilde{k},
\end{equation}
where we have defined

\begin{equation}\label{eq:INT:kt}
 \tilde{k}=\frac{\pi N^2 R^2}{2}k,
\end{equation}
in which $R$ is the fixed internuclear separation and $N$ is an arbitrary
normalization constant appearing as a factor in the explicit forms of
$\bar{S}$ and  $\bar{C}$. The explicit form of the coefficient of proportionality in (\ref{eq:INT:kt}) is specific to the choice of basis. For example, in the aforementioned case of short-range potential scattering \cite{Burke1977,BurkeJoachain1995} the coefficient is simply $N^2/2$, assuming the radial $S$ and $C$ functions are each associated with a common spherical harmonic for angle-function normalization. 

Since $\hat{H}$ is Hermitian, the $\{\chi_{i}\}$ are all square-integrable and each term in (\ref{eq:INT:trialwaveEXPL}) is real-valued, we note that

\numparts
\begin{eqnarray}\label{eq:INT:HERM2}
\langle\bar{S}\psi_{\mathrm{G}},\chi_{i}\rangle&=\langle\chi_{i},\bar{S}\psi_{\mathrm{G}}\rangle\quad\quad&\left(i=1,\dots,M\right),\\\label{eq:INT:HERM3}
\langle\bar{C}\psi_{\mathrm{G}},\chi_{i}\rangle&=\langle\chi_{i},\bar{C}\psi_{\mathrm{G}}\rangle\quad\quad&\left(i=1,\dots,M\right),\\\label{eq:INT:HERM5}
\langle\chi_{i},\chi_{j}\rangle&=\langle\chi_{j},\chi_{i}\rangle\quad\quad&\left(i,j=1,\dots,M\right).
\end{eqnarray}
\endnumparts
Henceforth, we will make implicit use of these properties.

In the generalized Kohn method, a stationary principle is used to determine optimal values of the unknown parameters, $a_{\mathrm{t}}$ and $\{p_{1},\dots,p_{M}\}$, in \trialwave. This principle manifests itself in the system of linear equations,

\begin{equation}\label{eq:KohnEq}
 Ax=-b,
\end{equation}
where

\numparts
\begin{eqnarray}\label{eq:matA}
A&=&\left[ \begin{array}{cccc} 
\langle\bar{C}\psi_{\mathrm{G}},\bar{C}\psi_{\mathrm{G}}\rangle & \langle\bar{C}\psi_{\mathrm{G}},\chi_{1}\rangle &
\cdots& 
\langle\bar{C}\psi_{\mathrm{G}},\chi_{M}\rangle \\  
\langle\chi_{1},\bar{C}\psi_{\mathrm{G}}\rangle & \langle\chi_{1},\chi_{1}\rangle &
\cdots &
\langle\chi_{1},\chi_{M}\rangle \\
\vdots & \vdots & \ddots & \vdots \\
 \langle\chi_{M},\bar{C}\psi_{\mathrm{G}}\rangle & \langle\chi_{M},\chi_{1}\rangle &
\cdots & 
\langle\chi_{M},\chi_{M}\rangle
\end{array}\right],\\
 \label{eq:vecB}b&=&\left[\begin{array}{c}
          \langle\bar{C}\psi_{\mathrm{G}},\bar{S}\psi_{\mathrm{G}}\rangle \\
	 \langle\chi_{1},\bar{S}\psi_{\mathrm{G}}\rangle \\
	 \vdots \\
	 \langle\chi_{M},\bar{S}\psi_{\mathrm{G}}\rangle \\
         \end{array}\right],\\ 
x&=&\left[\begin{array}{c}
          a_{t}\\
	  p_{1}\\
	  \vdots\\
 	  p_{M}
          \end{array}\right]. \label{eq:matAc}
\end{eqnarray}
\endnumparts
If $A$ is nonsingular then the solution of (\ref{eq:KohnEq}) uniquely
determines optimal values for the unknown parameters in \trialwave. This
solution can then be used to calculate a variational approximation,
\ps$\in\left(-\pi/2,\pi/2\right]$, to the exact scattering phase shift,
$\eta$. We note for generality that if the method is applied to single channel
electron scattering with exchange, whilst the definition of
$\langle X,Y\rangle$ changes to account for antisymmetrization,
the abstract forms of (\ref{eq:INT:sccs}) and (\ref{eq:INT:HERM2})--(\ref{eq:matAc}) are unaltered. 

For $S$ and $C$ as in
\cite{CooperArmourPlummer2009}, \ps is obtained implicitly from the
definition,

\begin{equation}\label{eq:INT:tan}
\tan\left(\eta_{\mathrm{v}}-\tau+c\right)=a_{\mathrm{t}}-\frac{1}{\tilde{k}}\mathcal{I}\left[\Psi_{\mathrm{t}}\right],
\end{equation}
in which $c=kR/2$. The appearance of the parameter, $c$, in
(\ref{eq:INT:tan}) is an artefact of the particular choice of basis functions,
$S$ and $C$, in our \ehmol calculations. When using prolate spheroidal coordinates, it is convenient to construct these functions in such a way that they include a phase factor equal to $-c$. The appearance of $c$ in (\ref{eq:INT:tan}) merely compensates for this phase factor and plays no `active' part in the following analysis. 

The functional, $\mathcal{I}\left[\Psi_{\mathrm{t}}\right]$, is defined as 

\begin{equation}\label{eq:INT:IFunc}
 \mathcal{I}\left[\Psi_{\mathrm{t}}\right] = \langle \Psi_{\mathrm{t}},\Psi_{\mathrm{t}}\rangle 
\end{equation}
and, when the Kohn equations (\ref{eq:KohnEq}) are satisfied, can be shown to take the form 

\begin{equation}\label{eq:APT:redI}
\mathcal{I}\left[\Psi_{\mathrm{t}}\right]=\langle\bar{S}\psi_{\mathrm{G}},\bar{S}\psi_{\mathrm{G}}\rangle+a_{\mathrm{t}}\langle\bar{S}\psi_{\mathrm{G}},\bar{C}\psi_{\mathrm{G}}\rangle+\sum_{j=1}^{M} p_{j}\langle\bar{S}\psi_{\mathrm{G}},\chi_{j}\rangle.
\end{equation}
Under these circumstances, upon substitution of the solution of
(\ref{eq:KohnEq}) into (\ref{eq:INT:tan}), the error in
$\tan\left(\eta_{\mathrm{v}}-\tau+c\right)$ from
$\tan\left(\eta-\tau+c\right)$ can be shown to be second order in the error of
\trialwave from $\Psi$ \cite{Kohn1948,Kato1950,Kato1951}. This is the attraction of the Kohn variational
principle.

\subsection{Singularities and anomalous behaviour}\label{ss:anom}

In the case where $A$ is singular, the system of Kohn equations (\ref{eq:KohnEq}) either has no unique solution or no solution at all, and the variational method breaks down. Where $A$ is nonsingular, it has been widely documented \cite{CooperArmourPlummer2009,Schwartz1961,Schwartz1961b,Nesbet1968,Brownstein1968,Shimamura1971,Nesbet1978,Takatsuka1979} that the results of Kohn calculations can exhibit nonphysical behaviour when the values of various parameters in the Kohn trial function are close to values making $A$ singular. In the generalized Kohn method, problems associated with these so-called Schwartz singularities are accounted for by the inclusion of the adjustable phase parameter, $\tau\in\left[0,\pi\right)$. As we have already noted \cite{CooperArmourPlummer2009}, for a sufficiently accurate trial function the choice of $\tau$ should have no significant physical effect on the calculation of \ps and, in this respect, we can regard it as a free parameter. Consequently, if $\Delta\eta_{\mathrm{v}}$ denotes the difference in the phase shift approximations obtained at two similar values of $\tau$ separated by $\Delta\tau$, the typical dependence of \ps on $\tau$ is such that $\Delta\eta_{\mathrm{v}}/\Delta\tau$ is small. However, if $\tau$ is close to a value, $\tau_{\mathrm{s}}$, making $A$ singular, then nonphysical anomalies can arise and these are characterized by large values of $\Delta\eta_{\mathrm{v}}/\Delta\tau$. We have found \cite{CooperArmourPlummer2009} that a choice of $\tau\in\left[0,\pi\right)$ can normally be made successfully to avoid anomalies of this kind at a given positron energy.

We shall now develop a framework formally to describe the underlying mathematical structure of those anomalies in \pst encountered by varying $\tau$ at a fixed $k$. This framework offers a complete, analytic description of \pst. We begin by considering the determinant of $A$ (\ref{eq:matA}); from the Laplace expansion of \deter along the first row of $A$ , then subsequently along the first column of each resulting submatrix of $A$, using (\ref{eq:tautrans}) we obtain

\begin{eqnarray}
\nonumber\det\left(A\right)&=&\mathcal{P}\left(k\right)\langle\bar{C}\psi_{\mathrm{G}},\bar{C}\psi_{\mathrm{G}}\rangle+\sum_{i=1}^{M}\sum_{j=1}^{M}\mathcal{S}_{ij}\left(k\right)\langle\bar{C}\psi_{\mathrm{G}},\chi_{i}\rangle\langle \bar{C}\psi_{\mathrm{G}},\chi_{j}\rangle\\\label{eq:COMP:determinant}
&=&\mathcal{A}\left(k\right)\sin^{2}\left(\tau\right)+\mathcal{B}\left(k\right)\sin\left(\tau\right)\cos\left(\tau\right)+\mathcal{C}\left(k\right)\cos^{2}\left(\tau\right),
\end{eqnarray}
where $\mathcal{A}\left(k\right)$, $\mathcal{B}\left(k\right)$, $\mathcal{C}\left(k\right)$, $\mathcal{P}\left(k\right)$ and $\mathcal{S}_{ij}\left(k\right)=\mathcal{S}_{ji}\left(k\right)$ depend on $k$ but are independent of $\tau$. If, by accident, $k=k_{\mathrm{z}}$ such that $\mathcal{A}\left(k_{\mathrm{z}}\right)=\mathcal{B}\left(k_{\mathrm{z}}\right)=\mathcal{C}\left(k_{\mathrm{z}}\right)=0$, then $\det\left(A;k=k_{\mathrm{z}}\right)$ is identically zero and a unique value of \ps cannot be determined from the Kohn equations at any value of $\tau$. We define 

\begin{equation}
Z=\lbrace k_{\mathrm{z}}\in\mathbb{R}^{+}:\mathcal{A}\left(k_{\mathrm{z}}\right)=\mathcal{B}\left(k_{\mathrm{z}}\right)=\mathcal{C}\left(k_{\mathrm{z}}\right)=0\rbrace.
\end{equation}
To ensure that our Kohn calculations can always be implemented consistently for some choice of $\tau$, we will henceforth consider only values of $k\in\mathbb{R}^{+}\setminus Z$ rather than $k\in\mathbb{R}^{+}$, although for brevity we will not usually mention this distinction explicitly. In practice, we have not yet found any such value of $k_{\mathrm{z}}$ in our numerical calculations. It is conceivable, though very unlikely, that such a value might occur by accident for some configuration of parameters in the trial wavefunction, but then it should be possible to vary the nonlinear parameters in the short-range correlation functions always to avoid this eventuality. Hence, we consider the case $k=k_{\mathrm{z}}$ here purely so that it can be excluded, in the interests of good mathematical practice. For each $k\in\mathbb{R}^{+}\setminus Z$, there exists at least one value of $\tau$ such that $A$ is nonsingular.

Next, consider the matrix, $\tilde{A}$, formed by replacing the first column of $A$ by $-b$, where $b$ is as in (\ref{eq:vecB}), so that

\begin{eqnarray}\label{eq:COMP:matAt}
\tilde{A}&=&\left[ \begin{array}{cccc} 
-\langle\bar{C}\psi_{\mathrm{G}},\bar{S}\psi_{\mathrm{G}}\rangle & \langle\bar{C}\psi_{\mathrm{G}},\chi_{1}\rangle &
\cdots& 
\langle\bar{C}\psi_{\mathrm{G}},\chi_{M}\rangle \\ 
-\langle\chi_{1},\bar{S}\psi_{\mathrm{G}}\rangle & \langle\chi_{1},\chi_{1}\rangle &
\cdots &
\langle\chi_{1},\chi_{M}\rangle \\
\vdots & \vdots & \ddots & \vdots \\
 -\langle\chi_{M},\bar{S}\psi_{\mathrm{G}}\rangle & \langle\chi_{M},\chi_{1}\rangle &
\cdots & 
\langle\chi_{M},\chi_{M}\rangle
\end{array}\right].
\end{eqnarray}
By the same method used to find (\ref{eq:COMP:determinant}), we have

\begin{eqnarray}\label{eq:COMP:dett}
 \nonumber\det\left(\tilde{A}\right)&=&-\mathcal{P}\left(k\right)\langle\bar{C}\psi_{\mathrm{G}},\bar{S}\psi_{\mathrm{G}}\rangle
-\sum_{i=1}^{M}\sum_{j=1}^{M}\mathcal{S}_{ij}\left(k\right)\langle\bar{C}\psi_{\mathrm{G}},\chi_{i}\rangle\langle \bar{S}\psi_{\mathrm{G}},\chi_{j}\rangle\\
&=&\tilde{\mathcal{A}}\left(k\right)\sin^{2}\left(\tau\right)+\tilde{\mathcal{B}}\left(k\right)\sin\left(\tau\right)\cos\left(\tau\right)+\tilde{\mathcal{C}}\left(k\right)\cos^{2}\left(\tau\right),
\end{eqnarray}
where $\tilde{\mathcal{A}}\left(k\right)$, $\tilde{\mathcal{B}}\left(k\right)$ and $\tilde{\mathcal{C}}\left(k\right)$ depend on $k$ but are independent of $\tau$. The functions, $\mathcal{P}\left(k\right)$ and $\mathcal{S}_{ij}\left(k\right)$, are as in (\ref{eq:COMP:determinant}). Now, setting $\tau=0$, $\tau=\pi/2$ and $\tau=\pi/4$ successively in both (\ref{eq:COMP:determinant}) and (\ref{eq:COMP:dett}), it is readily shown that

\numparts
\begin{eqnarray}\label{eq:COMP:detzero}\fl
\mathcal{C}\left(k\right)&=&\mathcal{P}\left(k\right)\langle C\psi_{\mathrm{G}},C\psi_{\mathrm{G}}\rangle
+\sum_{i=1}^{M}\sum_{j=1}^{M}\mathcal{S}_{ij}\left(k\right)\langle C\psi_{\mathrm{G}},\chi_{i}\rangle\langle C\psi_{\mathrm{G}},\chi_{j}\rangle,\\\label{eq:COMP:dettzero}\fl
 \tilde{\mathcal{C}}\left(k\right)&=&-\mathcal{P}\left(k\right)\langle C\psi_{\mathrm{G}},S\psi_{\mathrm{G}}\rangle-\sum_{i=1}^{M}\sum_{j=1}^{M}\mathcal{S}_{ij}\left(k\right)\langle C\psi_{\mathrm{G}},\chi_{i}\rangle\langle S\psi_{\mathrm{G}}, \chi_{j}\rangle,\\\label{eq:COMP:detpi}\fl
 \mathcal{A}\left(k\right)&=&\mathcal{P}\left(k\right)\langle S\psi_{\mathrm{G}},S\psi_{\mathrm{G}}\rangle+\sum_{i=1}^{M}\sum_{j=1}^{M}\mathcal{S}_{ij}\left(k\right)\langle S\psi_{\mathrm{G}},\chi_{i}\rangle\langle S\psi_{\mathrm{G}}, \chi_{j}\rangle,\\\label{eq:COMP:dettpi}\fl
\tilde{\mathcal{A}}\left(k\right)&=&\mathcal{P}\left(k\right)\langle S\psi_{\mathrm{G}},C\psi_{\mathrm{G}}\rangle+\sum_{i=1}^{M}\sum_{j=1}^{M}\mathcal{S}_{ij}\left(k\right)\langle S\psi_{\mathrm{G}},\chi_{i}\rangle\langle C\psi_{\mathrm{G}},\chi_{j}\rangle,\\\nonumber\fl\mathcal{B}\left(k\right)&=&-\mathcal{P}\left(k\right)\left(\langle S\psi_{\mathrm{G}},C\psi_{\mathrm{G}}\rangle+\langle C\psi_{\mathrm{G}},S\psi_{\mathrm{G}}\rangle\right)\\\label{eq:COMP:detpi4}\fl&-&\sum_{i=1}^{M}\sum_{j=1}^{M}\mathcal{S}_{ij}\left(k\right)\left(\langle C\psi_{\mathrm{G}},\chi_{i}\rangle\langle S\psi_{\mathrm{G}}, \chi_{j}\rangle+\langle S\psi_{\mathrm{G}},\chi_{i}\rangle\langle C\psi_{\mathrm{G}},\chi_{j}\rangle\right),\\\nonumber\fl\tilde{\mathcal{B}}\left(k\right)&=&\mathcal{P}\left(k\right)\left(\langle S\psi_{\mathrm{G}},S\psi_{\mathrm{G}}\rangle-\langle C\psi_{\mathrm{G}},C\psi_{\mathrm{G}}\rangle\right)\\\label{eq:COMP:dettpi4}\fl
&+&\sum_{i=1}^{M}\sum_{j=1}^{M}\mathcal{S}_{ij}\left(k\right)\left(\langle S\psi_{\mathrm{G}},\chi_{i}\rangle\langle S\psi_{\mathrm{G}}, \chi_{j}\rangle-\langle C\psi_{\mathrm{G}},\chi_{i}\rangle\langle C\psi_{\mathrm{G}},\chi_{j}\rangle\right),
\end{eqnarray}
\endnumparts
from which we obtain the relations, 
\numparts
\begin{eqnarray}
 \mathcal{A}\left(k\right)-\mathcal{C}\left(k\right)&=&\tilde{\mathcal{B}}\left(k\right),\\
 \tilde{\mathcal{A}}\left(k\right)-\tilde{\mathcal{C}}\left(k\right)&=&-\mathcal{B}\left(k\right),
\end{eqnarray}
so that we may rewrite (\ref{eq:COMP:dett}) as 
\endnumparts
\begin{eqnarray}
\nonumber\det\left(\tilde{A}\right)&=&\tilde{\mathcal{A}}\left(k\right)\sin^{2}\left(\tau\right)+\left[\mathcal{A}\left(k\right)-\mathcal{C}\left(k\right)\right]\sin\left(\tau\right)\cos\left(\tau\right)\\\label{eq:COMP:dett2}
&+&\left[\tilde{\mathcal{A}}\left(k\right)+\mathcal{B}\left(k\right)\right]\cos^{2}\left(\tau\right).
\end{eqnarray}

Now, for nonsingular $A$, we define $\Theta\left(k\right)$ to be

\begin{equation}\label{eq:GEN:theta}
\Theta\left(k\right)=\det\left(A\right)\mathcal{I}\left[\Psi_{\mathrm{t}}\right].
\end{equation}
Although $\det\left(A\right)$ and $\mathcal{I}\left[\Psi_{\mathrm{t}}\right]$ each depend on $\tau$, with effort it is possible to show that the product of these two terms does not. A proof of this result is given in \ref{app:theta}. Next, by Cramer's rule, we can write

\begin{equation}\label{eq:Cramer}
 a_{\mathrm{t}}=\frac{\det\left(\tilde{A}\right)}{\det\left(A\right)},
\end{equation}
for nonsingular $A$. Using (\ref{eq:GEN:theta}) and (\ref{eq:Cramer}), for nonsingular $A$ we can then rewrite (\ref{eq:INT:tan}) as

\begin{equation}\label{eq:COMP:tan2}
 \tan\left(\eta_{\mathrm{v}}-\tau+c\right) = \frac{\det\left(\tilde{A}\right)-\Gamma\left(k\right)}{\det\left(A\right)},
\end{equation}
where we have defined the function, $\Gamma\left(k\right)$, which is independent of $\tau$, as

\begin{equation}\label{eq:COMP:gam}
 \Gamma\left(k\right)=\frac{\Theta\left(k\right)}{\tilde{k}}.
\end{equation}
Since, for each $k\in\mathbb{R}^{+}\setminus Z$, it is always possible to find a nonsingular $A$ for some choice of $\tau$, $\Theta\left(k\right)$ and $\Gamma\left(k\right)$ are defined completely in this domain. Further, for $S$ and $C$ as in \cite{CooperArmourPlummer2009}, both $\Theta\left(k\right)$ and $\Gamma\left(k\right)$ are continuous functions of $k$ over $\mathbb{R}^{+}\setminus Z$. 

We will find it convenient to define the functions,

\begin{eqnarray}
 \nonumber f\left(\tau;k\right)&=&\det\left(\tilde{A}\right)-\Gamma\left(k\right)\\\nonumber
&=&\tilde{\mathcal{A}}\left(k\right)\sin^{2}\left(\tau\right)+\left[\mathcal{A}\left(k\right)-\mathcal{C}\left(k\right)\right]\sin\left(\tau\right)\cos\left(\tau\right)\\\label{eq:COMP:f}
&+&\left[\tilde{\mathcal{A}}\left(k\right)+\mathcal{B}\left(k\right)\right]\cos^{2}\left(\tau\right)-\Gamma\left(k\right)
\end{eqnarray}
and

\begin{equation}\label{eq:COMP:g}\fl
g\left(\tau;k\right)=\det\left(A\right)=\mathcal{A}\left(k\right)\sin^{2}\left(\tau\right)+\mathcal{B}\left(k\right)\sin\left(\tau\right)\cos\left(\tau\right)+\mathcal{C}\left(k\right)\cos^{2}\left(\tau\right),
\end{equation}
where we have explicitly denoted the parametric dependence of $f\left(\tau;k\right)$ and $g\left(\tau;k\right)$ on $k$. For nonsingular $A$, we then have

\begin{equation}\label{eq:COMP:tan3}
 \tan\left(\eta_{\mathrm{v}}-\tau+c\right) = \frac{f\left(\tau;k\right)}{g\left(\tau;k\right)}.
\end{equation}
Next, we see that

\begin{equation}\label{eq:COMP:fp}
f'\left(\tau;k\right)=\left[\mathcal{A}\left(k\right)-\mathcal{C}\left(k\right)\right]\cos\left(2\tau\right)-\mathcal{B}\left(k\right)\sin\left(2\tau\right)
\end{equation}
and 

\begin{equation}\label{eq:COMP:gp}
g'\left(\tau;k\right)=\left[\mathcal{A}\left(k\right)-\mathcal{C}\left(k\right)\right]\sin\left(2\tau\right)+\mathcal{B}\left(k\right)\cos\left(2\tau\right),
\end{equation}
where, in both cases, the prime indicates partial differentiation with respect to $\tau$. Moreover, differentiating (\ref{eq:COMP:tan3}) with respect to $\tau$, we find

\begin{eqnarray}\nonumber\fl
 \sec^{2}\left(\eta_{\mathrm{v}}-\tau+c\right)\left(\frac{\partial\eta_{\mathrm{v}}}{\partial\tau}-1\right) &=& \left[1+\frac{f^2\left(\tau;k\right)}{g^2\left(\tau;k\right)}\right]\left[\frac{\partial\eta_{\mathrm{v}}}{\partial\tau}-1\right]\\\fl
&=&\frac{g\left(\tau;k\right) f'\left(\tau;k\right)-f\left(\tau;k\right) g'\left(\tau;k\right)}{g^2\left(\tau;k\right)},
\end{eqnarray}
so that

\begin{equation}\label{eq:COMP:derivfg}
 \frac{\partial\eta_{\mathrm{v}}}{\partial\tau} = \frac{f^2\left(\tau;k\right)+g^2\left(\tau;k\right)+g\left(\tau;k\right)f'\left(\tau;k\right)-f\left(\tau;k\right)g'\left(\tau;k\right)}{f^2\left(\tau;k\right)+g^2\left(\tau;k\right)}.
\end{equation}
It can be shown using (\ref{eq:COMP:f}), (\ref{eq:COMP:g}), (\ref{eq:COMP:fp}) and (\ref{eq:COMP:gp}) that

\begin{eqnarray}\nonumber
 &&f^2\left(\tau;k\right)+g^2\left(\tau;k\right)+g\left(\tau;k\right)f'\left(\tau;k\right)-f\left(\tau;k\right)g'\left(\tau;k\right)\\ =&&\left[\tilde{\mathcal{A}}\left(k\right)-\Gamma\left(k\right)\right]\left[\tilde{\mathcal{A}}\left(k\right)-\Gamma\left(k\right)+\mathcal{B}\left(k\right)\right]+\mathcal{A}\left(k\right)\mathcal{C}\left(k\right),
\end{eqnarray}
which is independent of $\tau$. We define 

\begin{equation}\label{eq:FORM:calG}
\mathcal{G}\left(k\right)=\left[\tilde{\mathcal{A}}\left(k\right)-\Gamma\left(k\right)\right]\left[\tilde{\mathcal{A}}\left(k\right)-\Gamma\left(k\right)+\mathcal{B}\left(k\right)\right]+\mathcal{A}\left(k\right)\mathcal{C}\left(k\right)
\end{equation}
and, further, prove in \ref{app:gam} that 

\begin{equation}\label{eq:FORM:GAMAGO}
\mathcal{G}\left(k\right)=\Gamma^{2}\left(k\right).
\end{equation}
It follows immediately that
\begin{equation}\label{eq:FORM:finalderiv}
 \frac{\partial\eta_{\mathrm{v}}}{\partial\tau} = \frac{\Gamma^{2}\left(k\right)}{f^2\left(\tau;k\right)+g^2\left(\tau;k\right)},
\end{equation}
for nonsingular $A$. 

Anomalous behaviour in \pst will arise whenever $\tau$ is such that the ratio given by (\ref{eq:FORM:finalderiv}) is unusually large. Singularities in a Kohn calculation appear whenever $g\left(\tau;k\right)=0$, though it is clear from (\ref{eq:FORM:finalderiv}) that the presence of singularities is neither a sufficient nor a necessary condition for anomalies to occur. For a fixed value of $k$, values of \fderiv might be large even if $g\left(\tau;k\right)$ is never zero. Conversely, for fixed $k$, large values of \fderiv are not guaranteed in the limit as $\tau$ varies so that $g\left(\tau;k\right)\rightarrow0$, owing to the fact that both $f\left(\tau;k\right)$ and $g\left(\tau;k\right)$ appear in the denominator of the expression for \fderiv and their zeros will not coincide in general. This is the mathematical origin of the anomaly-free singularities reported previously \cite{CooperArmourPlummer2009}; in our own calculations, we have confirmed that those singularities are of such a type that, in the limit as $\tau$ varies so that $g\left(\tau;k\right)\rightarrow0$, $f^{2}\left(\tau;k\right)$ does not become small in comparison to $\Gamma^2\left(k\right)$.

An immediate consequence of (\ref{eq:FORM:GAMAGO}) is that, using (\ref{eq:FORM:calG}) and provided $k$ is such that $2\tilde{\mathcal{A}}\left(k\right)+\mathcal{B}\left(k\right)\neq0$, we can write

\begin{equation}\label{eq:FORM:gsim}
\Gamma\left(k\right)=\frac{\left[\tilde{\mathcal{A}}\left(k\right)+\mathcal{B}\left(k\right)\right]\tilde{\mathcal{A}}\left(k\right)+\mathcal{A}\left(k\right)\mathcal{C}\left(k\right)}{2\tilde{\mathcal{A}}\left(k\right)+\mathcal{B}\left(k\right)},
\end{equation}
so that the functions of $k$ in (\ref{eq:COMP:f}) can be expressed purely in terms of $\mathcal{A}\left(k\right)$, $\mathcal{B}\left(k\right)$, $\mathcal{C}\left(k\right)$ and $\tilde{\mathcal{A}}\left(k\right)$. Hence, in calculating the value of \ps from (\ref{eq:COMP:tan3}), in general there is actually no need to solve the Kohn equations (\ref{eq:KohnEq}). All that is required to determine \pst completely at each $k$ is to find the values of the four determinants, $\mathcal{A}\left(k\right)$, $\mathcal{B}\left(k\right)$, $\mathcal{C}\left(k\right)$ and $\tilde{\mathcal{A}}\left(k\right)$; substituting (\ref{eq:FORM:gsim}) into (\ref{eq:COMP:f}), we find that (\ref{eq:COMP:tan3}) can be rewritten

\begin{equation}\label{eq:FORM:NOSOL}\fl
\tan\left(\eta_{\mathrm{v}}-\tau+c\right)=\frac{\left[\mathcal{A}\left(k\right)-\mathcal{C}\left(k\right)\right]\sin\left(\tau\right)\cos\left(\tau\right)+\mathcal{B}\left(k\right)\cos^{2}\left(\tau\right)+\mathcal{D}\left(k\right)}{\mathcal{A}\left(k\right)\sin^{2}\left(\tau\right)+\mathcal{B}\left(k\right)\sin\left(\tau\right)\cos\left(\tau\right)+\mathcal{C}\left(k\right)\cos^{2}\left(\tau\right)},
\end{equation}
for nonsingular $A$, where we have defined

\begin{equation}\label{eq:FORM:calD}
 \mathcal{D}\left(k\right)=\frac{\tilde{\mathcal{A}}^{2}\left(k\right)-\mathcal{A}\left(k\right)\mathcal{C}\left(k\right)}{2\tilde{\mathcal{A}}\left(k\right)+\mathcal{B}\left(k\right)}.
\end{equation}
If $k=k_{\mathrm{s}}\in\mathbb{R}^{+}\setminus Z$ such that
$2\tilde{\mathcal{A}}\left(k_{\mathrm{s}}\right)+\mathcal{B}\left(k_{\mathrm{s}}\right)=0$,
then from (\ref{eq:APG:apr}) and (\ref{eq:APG:td}) we also have
$\left[\tilde{\mathcal{A}}\left(k_{\mathrm{s}}\right)+\mathcal{B}\left(k_{\mathrm{s}}\right)\right]\tilde{\mathcal{A}}\left(k_{\mathrm{s}}\right)+\mathcal{A}\left(k_{\mathrm{s}}\right)\mathcal{C}\left(k_{\mathrm{s}}\right)=0$.
Here, the continuity of $\Gamma\left(k\right)$ over
$\mathbb{R}^{+}\setminus Z$ ensures that the value of
$\Gamma\left(k\right)$ determined from (\ref{eq:FORM:gsim}) in the limit as
$k\rightarrow k_{\mathrm{s}}$ is well defined and equal to the value of
$\Gamma\left(k_{\mathrm{s}}\right)$ determined using (\ref{eq:COMP:gam}),
(\ref{eq:APT:A5}), (\ref{eq:APT:A13}), (\ref{eq:APT:A14}) and (\ref{eq:APT:finalresult}). In practical
calculations, $\Gamma\left(k_{\mathrm{s}}\right)$ may be determined either by this
method or by interpolation of $\Gamma\left(k\right)$ either side of
$k_{\mathrm{s}}$. The physical significance of instances of $k_{\mathrm{s}}$ is discussed in
section \ref{sec:familiar}.

\subsection{Optimization}\label{ss:opt}

Having found an analytical form for \fderiv, an obvious extension to our investigation is to optimize this expression with respect to $\tau$. From inspection of (\ref{eq:FORM:finalderiv}) we have $\partial\eta_{\mathrm{v}}/\partial\tau\geq0$ so, at each $k$, finding a global minimum of \fderiv with respect to $\tau\in\left[0,\pi\right)$ locates the point at which \pst varies most slowly with $\tau$. This forms a natural optimization scheme for choosing $\tau$ to avoid anomalous behaviour.

For nonsingular $A$, partial differentiation of (\ref{eq:FORM:finalderiv}) with respect to $\tau$ gives

\begin{equation}\label{eq:FORM:secondderiv}
\frac{\partial^{2}\eta_{\mathrm{v}}}{\partial\tau^{2}}=-2\frac{\Gamma^{2}\left(k\right)}{\left[f^2\left(\tau;k\right)+g^2\left(\tau;k\right)\right]^2}\left[f\left(\tau;k\right)f'\left(\tau;k\right)+g\left(\tau;k\right)g'\left(\tau;k\right)\right]
\end{equation}
and, after some manipulation, we find

\begin{equation}\label{eq:FORM:xyeqnz}
f\left(\tau;k\right)f'\left(\tau;k\right)+g\left(\tau;k\right)g'\left(\tau;k\right)=\mathcal{X}\left(k\right)\sin\left(2\tau\right)+\mathcal{Y}\left(k\right)\cos\left(2\tau\right),
\end{equation}
where we have defined 

\begin{equation}\label{eq:FORM:x}
\mathcal{X}\left(k\right)=\frac{\mathcal{A}^2\left(k\right)-\mathcal{B}^2\left(k\right)-\mathcal{C}^2\left(k\right)}{2}+\mathcal{B}\left(k\right)\left[\Gamma\left(k\right)-\tilde{\mathcal{A}}\left(k\right)\right]
\end{equation}
and

\begin{equation}\label{eq:FORM:y}
\mathcal{Y}\left(k\right)=\left[\Gamma\left(k\right)-\tilde{\mathcal{A}}\left(k\right)\right]\left[\mathcal{C}\left(k\right)-\mathcal{A}\left(k\right)\right]+\mathcal{A}\left(k\right)\mathcal{B}\left(k\right),
\end{equation}
both of which are independent of $\tau$. Denoting by $\tau_{i}$ any value of $\tau$ for which $\partial^{2}\eta_{\mathrm{v}}/\partial\tau^{2}=0$, if $\Gamma\left(k\right)\neq0$ we then have

\begin{equation}\label{eq:FORM:xyeq}
 \mathcal{X}\left(k\right)\sin\left(2\tau_{i}\right)+\mathcal{Y}\left(k\right)\cos\left(2\tau_{i}\right)=0.
\end{equation}
In the special case where $k=k_{\mathrm{g}}$ such that $\Gamma\left(k_{\mathrm{g}}\right)=0$, then \fderiv is everywhere zero and optimization is not required since \ps is constant over $\tau$. In the special case where $k=k_{\mathrm{h}}$, such that $\mathcal{X}\left(k_{\mathrm{h}}\right)=\mathcal{Y}\left(k_{\mathrm{h}}\right)=0$, then for nonsingular $A$ the value of \sderiv is everywhere zero. Hence, \fderiv is constant with respect to variations in $\tau$. If this constant is equal to zero, optimization of \fderiv is not required since then \ps is constant over $\tau$. If this constant is nonzero, \ps varies linearly with $\tau$ and no preferred optimization can reasonably be defined. 

Discounting these two special cases, at each $k$ there will be exactly two values of $\tau_{i}\in\left[0,\pi\right)$ satisfying (\ref{eq:FORM:xyeq}), separated by $\pi/2$. In this case, using (\ref{eq:FORM:xyeqnz}) and (\ref{eq:FORM:xyeq}) and differentiating (\ref{eq:FORM:secondderiv}) with respect to $\tau$, at $\tau=\tau_{i}$ and for nonsingular $A$ we see that

\begin{equation}\label{eq:FORM:thirdderiv}
\frac{\partial^{3}\eta_{\mathrm{v}}}{\partial\tau^{3}}\left(\tau=\tau_{i}\right)=-4\Gamma^{2}\left(k\right)\frac{\left[\mathcal{X}\left(k\right)\cos\left(2\tau_{i}\right)-\mathcal{Y}\left(k\right)\sin\left(2\tau_{i}\right)\right]}{\left[f^2\left(\tau_{i};k\right)+g^2\left(\tau_{i};k\right)\right]^2}.
\end{equation}
In general, when $k$ is such that both $\Gamma\left(k\right)$ and at least one of $\mathcal{X}\left(k\right)$ or $\mathcal{Y}\left(k\right)$ is nonzero, then (\ref{eq:FORM:xyeq}) ensures that \tderiv is nonzero at $\tau=\tau_{i}$. Moreover, since the two values of $\tau_{i}$ are separated by $\pi/2$, we see from (\ref{eq:FORM:thirdderiv}) that the signs of \tderiv at the two values of $\tau_{i}$ are opposite. Hence, in general, \fderiv has one minimum and one maximum for $\tau\in\left[0,\pi\right)$. At each $k$, we will denote by $\tau_{0}$ and $\tau_{1}$ the values of $\tau_{i}$ respectively minimizing and maximizing \fderiv. We will denote the values of $\eta_{\mathrm{v}}\left(\tau=\tau_{0}\right)$ and $\eta_{\mathrm{v}}\left(\tau=\tau_{1}\right)$ respectively by $\eta_{\mathrm{v}}^{\left(0\right)}$ and $\eta_{\mathrm{v}}^{\left(1\right)}$.  

Next, assuming that $A$ is nonsingular and $k$ is such that $\tau_{0}$ and $\tau_{1}$ exist, we note the following. Firstly, we see from (\ref{eq:FORM:x}) and (\ref{eq:FORM:y}) that these conditions preclude having $k=k_{\mathrm{c}}$ such that both $\mathcal{A}\left(k_{\mathrm{c}}\right)=\mathcal{C}\left(k_{\mathrm{c}}\right)$ and $\mathcal{B}\left(k_{\mathrm{c}}\right)=0$, since $\mathcal{X}\left(k_{\mathrm{c}}\right)=\mathcal{Y}\left(k_{\mathrm{c}}\right)=0$ by inspection. Consequently, we note from (\ref{eq:COMP:fp}) and (\ref{eq:COMP:gp}) that $f'\left(\tau_{i};k\right)$ and $g'\left(\tau_{i};k\right)$ cannot both be zero. In fact, $f'\left(\tau_{i};k\right)$ cannot be zero, since this would require $g'\left(\tau_{i};k\right)\neq0$ and, using (\ref{eq:FORM:xyeq}), we see that (\ref{eq:FORM:xyeqnz}) could not then be satisfied, since we have assumed that $g\left(\tau_{i};k\right)\neq0$. Hence, using (\ref{eq:COMP:tan3}), (\ref{eq:FORM:xyeqnz}) and (\ref{eq:FORM:xyeq}), we write

\begin{equation}\label{eq:FORM:newprop}
f'\left(\tau_{i};k\right)\tan\left(\eta^{\left(i\right)}_{\mathrm{v}}-\tau_{i}+c\right)+g'\left(\tau_{i};k\right)=0,
\end{equation}
where $f'\left(\tau_{i};k\right)\neq0$.
Both $\tau_{0}$ and $\tau_{1}$ must satisfy (\ref{eq:FORM:newprop}). Since $\tau_{0}$ and $\tau_{1}$ are separated by $\pi/2$, using (\ref{eq:COMP:fp}) and (\ref{eq:COMP:gp}) we can then immediately conclude

\begin{equation}\label{eq:pssep}
\tan\left(\eta^{\left(0\right)}_{\mathrm{v}}-\tau_{0}+c\right)=\tan\left(\eta^{\left(1\right)}_{\mathrm{v}}-\tau_{1}+c\right),
\end{equation}
so that, for $\eta_{\mathrm{v}}\in\left(-\pi/2,\pi/2\right]$, the values of $\eta_{\mathrm{v}}^{\left(0\right)}$ and $\eta_{\mathrm{v}}^{\left(1\right)}$ must also be separated by $\pi/2$. In fact, using (\ref{eq:COMP:fp}), (\ref{eq:COMP:gp}), (\ref{eq:FORM:x}), (\ref{eq:FORM:y}), (\ref{eq:FORM:xyeq}) and (\ref{eq:FORM:newprop}), it is straightforward to show that

\begin{eqnarray}\nonumber
\tan\left(\eta^{\left(i\right)}_{\mathrm{v}}-\tau_{i}+c\right)&=&\frac{\left[\mathcal{A}\left(k\right)-\mathcal{C}\left(k\right)\right]\mathcal{Y}\left(k\right)-\mathcal{B}\left(k\right)\mathcal{X}\left(k\right)}{\left[\mathcal{A}\left(k\right)-\mathcal{C}\left(k\right)\right]\mathcal{X}\left(k\right)+\mathcal{B}\left(k\right)\mathcal{Y}\left(k\right)}\\\label{eq:FORM:tanpreal}&=&\frac{2\tilde{\mathcal{A}}\left(k\right)+\mathcal{B}\left(k\right)-2\Gamma\left(k\right)}{\mathcal{A}\left(k\right)+\mathcal{C}\left(k\right)}.
\end{eqnarray}

\section{The complex Kohn method}
\subsection{Variational approximations to the phase shift}

The complex Kohn method is an extension of the original variational approach
in which the boundary conditions of the trial wavefunction are
complex-valued. Although originally thought \cite{McCurdy1987,Schneider1988}
entirely to be free of Schwartz-type behaviour, anomalies have been reported
by Lucchese \cite{Lucchese1989} and, recently, by Cooper and coworkers
\cite{CooperArmourPlummer2009}. We now consider an implementation of the
complex Kohn method which is analogous to that described previously
\cite{CooperArmourPlummer2009}. Following the same approach as in the preceding
section, we now develop a mathematical formalism to explain the
success of the complex Kohn method in avoiding a particular class of anomalies.
The same notes and minor caveats concerning applications of the method to
systems other than \ehmol apply as in subsection \ref{sec:realintro}. 

We begin with a complex-valued trial wavefunction, viz.

\begin{equation}\label{eq:COMP:complextrialwave}
\breve{\Psi}_{\mathrm{t}} = \left(\bar{S} + a'_{\mathrm{t}}\bar{T}\right)\psi_{\mathrm{G}} + \sum_{i=1}^{M} p'_{i}\chi_{i},
\end{equation}
where

\begin{equation}\label{eq:COMP:hankel}
 \bar{T}=\bar{S}+\rmi\bar{C},
\end{equation}
the functions, $\bar{S}$, $\bar{C}$, \targ and $\{\chi_{1},\dots,\chi_{M}\}$ being as in (\ref{eq:INT:trialwaveEXPL}). The unknowns, $\{a'_{\mathrm{t}},p'_{1},\dots,p'_{M}\}$, will not, in general, be real. The primes on $\{a'_{\mathrm{t}},p'_{1},\dots,p'_{M}\}$ distinguish them from the corresponding values found in (\ref{eq:INT:trialwaveEXPL}). Throughout this section, unless otherwise noted we will use primes in this way to distinguish various quantities used in our application of the complex Kohn method from the corresponding quantities used in our application of the generalized Kohn method involving a real-valued trial function.

Application of the Kohn stationary principle leads to a set of equations analogous to (\ref{eq:KohnEq}),

\begin{equation}\label{eq:COMP:KohnEq}
 A'x'=-b',
\end{equation}
where

\numparts
\begin{eqnarray}\label{eq:COMP:matAP}
A'&=&\left[ \begin{array}{cccc} 
\langle\bar{T}^{*}\psi_{\mathrm{G}},\bar{T}\psi_{\mathrm{G}}\rangle & \langle\bar{T}^{*}\psi_{\mathrm{G}},\chi_{1}\rangle &
\cdots& 
\langle\bar{T}^{*}\psi_{\mathrm{G}},\chi_{M}\rangle \\ 
\langle\chi_{1},\bar{T}\psi_{\mathrm{G}}\rangle & \langle\chi_{1},\chi_{1}\rangle &
\cdots &
\langle\chi_{1},\chi_{M}\rangle \\
\vdots & \vdots & \ddots & \vdots \\
 \langle\chi_{M},\bar{T}\psi_{\mathrm{G}}\rangle & \langle\chi_{M},\chi_{1}\rangle &
\cdots & 
\langle\chi_{M},\chi_{M}\rangle
\end{array}\right],\\
 \label{eq:COMP:vecB}b'&=&\left[\begin{array}{c}
          \langle\bar{T}^{*}\psi_{\mathrm{G}},\bar{S}\psi_{\mathrm{G}}\rangle \\
	 \langle\chi_{1},\bar{S}\psi_{\mathrm{G}}\rangle \\
	 \vdots \\
	 \langle\chi_{M},\bar{S}\psi_{\mathrm{G}}\rangle \\
         \end{array}\right],\\\label{eq:COMP:solX}
x'&=&\left[\begin{array}{c}
          a_{t}'\\
	  p_{1}'\\
	  \vdots\\
 	  p_{M}'
          \end{array}\right].
\end{eqnarray}
\endnumparts
Here, $\bar{T}^{*}$ is the complex conjugate of $\bar{T}$. In the usual Dirac
notation, $\langle\bar{T}\vert$ implies complex conjugation of $\bar{T}$. However, as
pointed out by Chamberlain \cite{Chamberlain2002} (see also \cite{McCurdy1987,Schneider1988}), in a consistent
implementation of the complex Kohn method this conjugation of the `radial'
function should not, in fact, be performed. Hence, we have replaced $\langle\bar{T}\vert$ by $\langle\bar{T}^{*}\vert$ to indicate that the conjugation is not carried out.

If $A'$ is nonsingular then the solution of (\ref{eq:COMP:KohnEq}) uniquely determines optimal values for the unknown parameters in \ctrialwave. This solution can then be used to calculate a variational approximation, \psc$\in\mathbb{C}$, to the exact scattering phase shift. For $S$ and $C$ as in \cite{CooperArmourPlummer2009}, this estimate is obtained implicitly from the definition,

\begin{equation}\label{eq:COMP:tanps}
 \tan\left(\eta'_{\mathrm{v}}-\tau+c\right)=\frac{\rmi a'_{\mathrm{t}}-\tilde{k}^{-1}\mathcal{I}'\left[\breve{\Psi}_{\mathrm{t}}\right] }{1+a'_{\mathrm{t}}+\rmi\tilde{k}^{-1}\mathcal{I}'\left[\breve{\Psi}_{\mathrm{t}}\right]},
\end{equation}
where

\begin{equation}\label{eq:COMP:cIFunc}
\mathcal{I}'\left[\breve{\Psi}_{\mathrm{t}}\right]=\langle\breve{\Psi}^{*}_{\mathrm{t}}\vert\left(\hat{H}-E\right)\vert\breve{\Psi}_{\mathrm{t}}\rangle=\langle\breve{\Psi}^{*}_{\mathrm{t}},\breve{\Psi}_{\mathrm{t}}\rangle
\end{equation}
is analogous to (\ref{eq:INT:IFunc}). In subsection \ref{ss:aab}, we demonstrate that the denominator of (\ref{eq:COMP:tanps}) is nonzero when both $A$ and $A'$ are nonsingular. When the Kohn equations (\ref{eq:COMP:KohnEq}) are satisfied, $\mathcal{I}'\left[\breve{\Psi}_{\mathrm{t}}\right]$ takes a form analogous to (\ref{eq:APT:redI}),
 
\begin{equation}\label{eq:COMP:redI}
\mathcal{I}'\left[\breve{\Psi}_{\mathrm{t}}\right]=\langle\bar{S}\psi_{\mathrm{G}},\bar{S}\psi_{\mathrm{G}}\rangle+a'_{\mathrm{t}}\langle\bar{S}\psi_{\mathrm{G}},\bar{T}\psi_{\mathrm{G}}\rangle+\sum_{j=1}^{M} p'_{j}\langle\bar{S}\psi_{\mathrm{G}},\chi_{j}\rangle.
\end{equation}
Under these circumstances, upon substitution of the solution of (\ref{eq:COMP:KohnEq}) into (\ref{eq:COMP:tanps}), the error in $\tan\left(\eta'_{\mathrm{v}}-\tau+c\right)$ from $\tan\left(\eta-\tau+c\right)$ can be shown to be second order in the error of \ctrialwave from $\Psi$.

It should be noted that the value of \psc will not, in general, be
real. However, since $\eta$ must be real, the imaginary part of \psc can
be regarded as an error term arising from the fact that the trial
function, \ctrialwave, is inexact. We will identify its precise form in subsection \ref{s:equiv}.

\subsection{Avoidance of anomalous behaviour}\label{ss:aab}

In the case where $A'$ is singular, the system of Kohn equations (\ref{eq:COMP:KohnEq}) either has no unique solution or no solution at all, and the variational method breaks down. For nonsingular $A'$, we now demonstrate that the value of \psc obtained in our implementation of the complex Kohn method is independent of the choice of $\tau$. We consider first the determinant of $A'$. Proceeding in a manner analogous to subsection \ref{ss:anom}, it is easily shown that

\begin{equation}\label{eq:COMP:detcfinal}
\det\left(A'\right)=\left[\mathcal{A}\left(k\right)-\mathcal{C}\left(k\right)-\rmi\mathcal{B}\left(k\right)\right]\exp\left(-2\rmi\tau\right),
\end{equation}
so that, as noted before \cite{CooperArmourPlummer2009}, \deterc describes a circle in the complex plane for variations of $\tau\in\left[0,\pi\right)$. Hence, in the complex Kohn method, singularities can neither be located nor avoided by varying only $\tau$. Here, $\mathcal{A}\left(k\right)$, $\mathcal{B}\left(k\right)$ and $\mathcal{C}\left(k\right)$ are as in (\ref{eq:COMP:determinant}). 

Next, consider the matrix, $\tilde{A}'$, formed by replacing the first column of $A'$ by $-b'$, so that

\begin{eqnarray}\label{eq:COMP:matAtP}
\tilde{A}'&=&\left[ \begin{array}{cccc} 
-\langle\bar{T}^{*}\psi_{\mathrm{G}},\bar{S}\psi_{\mathrm{G}}\rangle & \langle\bar{T}^{*}\psi_{\mathrm{G}},\chi_{1}\rangle &
\cdots& 
\langle\bar{T}^{*}\psi_{\mathrm{G}},\chi_{M}\rangle \\ 
-\langle\chi_{1},\bar{S}\psi_{\mathrm{G}}\rangle & \langle\chi_{1},\chi_{1}\rangle &
\cdots &
\langle\chi_{1},\chi_{M}\rangle \\
\vdots & \vdots & \ddots & \vdots \\
-\langle\chi_{M},\bar{S}\psi_{\mathrm{G}}\rangle & \langle\chi_{M},\chi_{1}\rangle &
\cdots & 
\langle\chi_{M},\chi_{M}\rangle
\end{array}\right].
\end{eqnarray}
Following the same approach taken in subsection \ref{ss:anom}, after a little work it is possible to show that

\begin{equation}\label{eq:COMP:dettpfin}\fl
\det\left(\tilde{A}'\right)=\left[\rmi\tilde{\mathcal{A}}\left(k\right)-\mathcal{C}\left(k\right)\right]+\left[\mathcal{C}\left(k\right)-\mathcal{A}\left(k\right)+\rmi\mathcal{B}\left(k\right)\right]\cos\left(\tau\right)\exp\left(-\rmi\tau\right),
\end{equation}
where $\mathcal{A}\left(k\right)$, $\mathcal{B}\left(k\right)$ and $\mathcal{C}\left(k\right)$ are as in (\ref{eq:COMP:determinant}) and $\tilde{\mathcal{A}}\left(k\right)$ is as in (\ref{eq:COMP:dett}). Next, we will find it convenient to define the functions

\begin{equation}\label{eq:COMP:u2}\fl
u\left(\tau;k\right)=-\rmi\mathcal{C}\left(k\right)-\tilde{\mathcal{A}}\left(k\right)+\left[\rmi\mathcal{C}\left(k\right)-\rmi\mathcal{A}\left(k\right)-\mathcal{B}\left(k\right)\right]\cos\left(\tau\right)\exp\left(-\rmi\tau\right)+\Gamma\left(k\right)
\end{equation}
and

\begin{equation}\label{eq:COMP:v2}
v\left(\tau;k\right)=\left[\mathcal{A}\left(k\right)-\mathcal{C}\left(k\right)-\rmi\mathcal{B}\left(k\right)\right]\exp\left(-2\rmi\tau\right)-\rmi u\left(\tau;k\right),
\end{equation}
noting that these functions satisfy the identity

\begin{equation}\label{eq:COMP:uvz}
u^2\left(\tau;k\right)+v^2\left(\tau;k\right)+v\left(\tau;k\right)u'\left(\tau;k\right)-u\left(\tau;k\right)v'\left(\tau;k\right)=0,
\end{equation}
where the primes on $u'\left(\tau;k\right)$ and $v'\left(\tau;k\right)$ indicate partial differentiation with respect to $\tau$. Next, in \ref{app:theta}, for nonsingular $A'$ we prove that 

\begin{equation}\label{eq:COMP:theta}
\det\left(A'\right)\mathcal{I}'\left[\breve{\Psi}_{\mathrm{t}}\right]=-\Theta\left(k\right),
\end{equation}
where $\Theta\left(k\right)$ is as defined in (\ref{eq:GEN:theta}). Using Cramer's rule, together with (\ref{eq:COMP:gam}), (\ref{eq:COMP:detcfinal}), (\ref{eq:COMP:dettpfin}), (\ref{eq:COMP:u2}), (\ref{eq:COMP:v2}) and (\ref{eq:COMP:theta}), we find that (\ref{eq:COMP:tanps}) can be rewritten

\begin{equation}\label{eq:COMP:fintan}
 \tan\left(\eta_{\mathrm{v}}'-\tau+c\right) = \frac{\rmi\det\left(\tilde{A}'\right)+\Gamma\left(k\right)}{\det\left(A'\right)+\det\left(\tilde{A}'\right)-\rmi\Gamma\left(k\right)}=\frac{u\left(\tau;k\right)}{v\left(\tau;k\right)},
\end{equation}
provided that $v\left(\tau;k\right)$ is nonzero. We note in passing that inspection of (\ref{eq:FORM:gsim}), (\ref{eq:COMP:detcfinal}), (\ref{eq:COMP:dettpfin}) and (\ref{eq:COMP:fintan}) indicates that the phase shift approximation in the complex Kohn method can be evaluated generally from the same four determinants required in the case of the generalized Kohn method and without the need to solve the Kohn equations (\ref{eq:COMP:KohnEq}).

By analogy with (\ref{eq:COMP:derivfg}), we then see that

\begin{equation}\label{eq:COMP:drat}
  \frac{\partial\eta_{\mathrm{v}}'}{\partial\tau} = \frac{u^2\left(\tau;k\right)+v^2\left(\tau;k\right)+v\left(\tau;k\right)u'\left(\tau;k\right)-u\left(\tau;k\right)v'\left(\tau;k\right)}{u^2\left(\tau;k\right)+v^2\left(\tau;k\right)}.
\end{equation}
Inspection of (\ref{eq:COMP:detcfinal}) and (\ref{eq:COMP:v2}) shows that the zeros of $u\left(\tau;k\right)$ and $v\left(\tau;k\right)$ coincide if and only if $A'$ is singular. Hence, for nonsingular $A'$, the denominator of (\ref{eq:COMP:drat}) is nonzero so that, using (\ref{eq:COMP:uvz}), (\ref{eq:COMP:drat}) becomes

\begin{equation}\label{eq:COMP:flat1}
 \frac{\partial\eta_{\mathrm{v}}'}{\partial\tau}=0,
\end{equation}
giving

\begin{equation}\label{eq:COMP:flat2}
   \frac{\partial\Re\left[\eta'_{\mathrm{v}}\right]}{\partial\tau}=\frac{\partial\Im\left[\eta'_{\mathrm{v}}\right]}{\partial\tau}=0.
\end{equation}
Thus, whenever \deterc and $v\left(\tau;k\right)$ are both nonzero, the value of \psc is independent of the choice of $\tau$ in (\ref{eq:COMP:complextrialwave}). Complex Kohn calculations of \psc will automatically be free of those Schwartz-type anomalies characterized by large values of \fderiv in the implementation of the generalized Kohn method already discussed in section \ref{s:genm}. Nevertheless, anomalies could still arise in the results of complex Kohn calculations due to the choice of some other parameter in the trial function; it is likely that this is the underlying cause of the persistent anomalies described earlier \cite{CooperArmourPlummer2009}.

If $v\left(\tau;k\right)=0$, then from (\ref{eq:COMP:uvz}) we either have $u\left(\tau;k\right)=v\left(\tau;k\right)=0$, in which case $\det\left(A'\right)=0$, or we have $u\left(\tau;k\right)=v'\left(\tau;k\right)$. We also find 

\begin{equation}
 \Im\left[v'\left(\tau;k\right)-u\left(\tau;k\right)\right]=\det\left(A\right),
\end{equation}
where $A$ is as in (\ref{eq:matA}). Hence, $v\left(\tau;k\right)$ is zero only if at least one of $A$ or $A'$ is singular. We will shortly carry out a formal comparison of the results of the generalized and complex Kohn methods. In so doing, we will take the parameters of our trial functions to be such that both $A$ and $A'$ are nonsingular, so that the Kohn equations in each case can uniquely be solved. These conditions automatically ensure that $v\left(\tau;k\right)\neq0$.

\subsection{Equivalence}\label{s:equiv}

We now demonstrate the effective equivalence of the generalized and complex Kohn variational methods. Consider the two trial wavefunctions, \trialwave and \ctrialwave, which contain the same approximate target wavefunction and identical sets of short-range correlation functions. Suppose that $A$ (\ref{eq:matA}) and $A'$ (\ref{eq:COMP:matAP}) are nonsingular so that solutions of the Kohn equations, (\ref{eq:KohnEq}) and (\ref{eq:COMP:KohnEq}), uniquely exist and we have $v\left(\tau;k\right)\neq0$. Suppose further that $k\neq k_{\mathrm{g}}$ and $k\neq k_{\mathrm{h}}$ in order that (\ref{eq:FORM:finalderiv}) can uniquely be minimized, as discussed in subsection \ref{ss:opt}. These conditions are sufficient for (\ref{eq:FORM:tanpreal}) and (\ref{eq:COMP:fintan}) to be well defined. Under these circumstances, we claim that

\begin{equation}\label{eq:COMP:toprove}
\Re\left[\eta'_{\mathrm{v}}\right]=\eta^{\left(0\right)}_{\mathrm{v}}.
\end{equation}
Here, we can regard $\Re\left[\eta'_{\mathrm{v}}\right]$ as the approximation to the phase shift in the complex Kohn method, since we have already noted that the imaginary part of \psc can be interpreted as an error term. We recall that $\eta^{\left(0\right)}_{\mathrm{v}}$ is that value of \ps obtained in the generalized Kohn method at the unique value, $\tau=\tau_{0}$, which minimizes \fderiv.

\begin{proof}
We consider $\tan\left(\eta_{\mathrm{v}}^{\left(i\right)}-\eta'_{\mathrm{v}}-\tau_{i}+\tau\right)$. Using (\ref{eq:COMP:f}), (\ref{eq:COMP:g}), (\ref{eq:FORM:tanpreal}), (\ref{eq:COMP:u2}), (\ref{eq:COMP:v2}) and (\ref{eq:COMP:fintan}), together with the standard result

\begin{equation}\label{eq:COMP:tandiff}
 \tan\left(P-Q\right)=\frac{\tan\left(P\right)-\tan\left(Q\right)}{1+\tan\left(P\right)\tan\left(Q\right)},
\end{equation}
after some considerable manipulation, it can be shown that

\begin{equation}\label{eq:COMP:retand}
\Re\left[\tan\left(\eta_{\mathrm{v}}^{\left(i\right)}-\eta'_{\mathrm{v}}-\tau_{i}+\tau\right)\right]=\frac{a\left(\tau;k\right)}{b\left(\tau;k\right)}
\end{equation}
and

\begin{equation}\label{eq:COMP:imtand}
\Im\left[\tan\left(\eta_{\mathrm{v}}^{\left(i\right)}-\eta'_{\mathrm{v}}-\tau_{i}+\tau\right)\right]=\frac{\Gamma^2\left(k\right)}{b\left(\tau;k\right)},
\end{equation}
where we have defined

\begin{equation}\label{eq:COMP:retantop}
a\left(\tau;k\right)=\mathcal{X}\left(k\right)\sin\left(2\tau\right)+\mathcal{Y}\left(k\right)\cos\left(2\tau\right)
\end{equation}
and

\begin{equation}\label{eq:COMP:retanbot}
b\left(\tau;k\right)=-f^{2}\left(\tau;k\right)-g^2\left(\tau;k\right),
\end{equation}
noting that $b\left(\tau;k\right)<0$, since we have assumed $g\left(\tau;k\right)\neq0$ so that $A$ is nonsingular. Derivations of (\ref{eq:COMP:retand}) and (\ref{eq:COMP:imtand}) are outlined in \ref{app:retan}. Now, setting $\tau=\tau_{i}$, we have $a\left(\tau_{i};k\right)=0$ from (\ref{eq:FORM:xyeq}). Hence,

\begin{equation}\label{eq:COMP:equiv}\fl
\Re\left[\tan\left(\eta_{\mathrm{v}}^{\left(i\right)}-\eta'_{\mathrm{v}}\right)\right]=\frac{\sin\left(\eta_{\mathrm{v}}^{\left(i\right)}-\Re\left[\eta'_{\mathrm{v}}\right]\right)\cos\left(\eta_{\mathrm{v}}^{\left(i\right)}-\Re\left[\eta'_{\mathrm{v}}\right]\right)}{\cos^2\left(\eta_{\mathrm{v}}^{\left(i\right)}-\Re\left[\eta'_{\mathrm{v}}\right]\right)+\sinh^2\left(\Im\left[\eta'_{\mathrm{v}}\right]\right)}=0
\end{equation}
and

\begin{equation}\label{eq:COMP:equiv2}\fl
\Im\left[\tan\left(\eta_{\mathrm{v}}^{\left(i\right)}-\eta'_{\mathrm{v}}\right)\right]=-\frac{\sinh\left(\Im\left[\eta'_{\mathrm{v}}\right]\right)\cosh\left(\Im\left[\eta'_{\mathrm{v}}\right]\right)}{\cos^2\left(\eta_{\mathrm{v}}^{\left(i\right)}-\Re\left[\eta'_{\mathrm{v}}\right]\right)+\sinh^2\left(\Im\left[\eta'_{\mathrm{v}}\right]\right)}=\frac{\Gamma^2\left(k\right)}{b\left(\tau_{i};k\right)}<0.
\end{equation}

Taking $\eta_{\mathrm{v}}^{\left(i\right)}\in\left(-\pi/2,\pi/2\right]$ and $\Re\left[\eta'_{\mathrm{v}}\right]\in\left(-\pi/2,\pi/2\right]$, since $\eta_{\mathrm{v}}^{\left(0\right)}$ and $\eta_{\mathrm{v}}^{\left(1\right)}$ are separated by $\pi/2$ we can immediately conclude from (\ref{eq:COMP:equiv}) that we have either $\Re\left[\eta'_{\mathrm{v}}\right]=\eta^{\left(0\right)}_{\mathrm{v}}$ or $\Re\left[\eta'_{\mathrm{v}}\right]=\eta^{\left(1\right)}_{\mathrm{v}}$. Moreover, using (\ref{eq:FORM:finalderiv}) and (\ref{eq:COMP:retanbot}), by the definitions of $\tau_{0}$ and $\tau_{1}$ it is plain that 

\begin{equation}\label{eq:COMP:ineq}
\frac{\Gamma^2\left(k\right)}{b\left(\tau_{1};k\right)}<\frac{\Gamma^2\left(k\right)}{b\left(\tau_{0};k\right)}<0,
\end{equation}
noting from (\ref{eq:COMP:equiv}) and (\ref{eq:COMP:equiv2}) that $b\left(\tau_{0};k\right)$ and $b\left(\tau_{1};k\right)$ cannot be equal, since $\eta_{\mathrm{v}}^{\left(0\right)}$ and $\eta_{\mathrm{v}}^{\left(1\right)}$ are separated by $\pi/2$ and exactly one of $\eta_{\mathrm{v}}^{\left(0\right)}$ or $\eta_{\mathrm{v}}^{\left(1\right)}$ must give $\cos\left(\eta_{\mathrm{v}}^{\left(i\right)}-\Re\left[\eta'_{\mathrm{v}}\right]\right)=0$. It follows directly from (\ref{eq:COMP:equiv2}) and (\ref{eq:COMP:ineq}) that $\cos^{2}\left(\eta_{\mathrm{v}}^{\left(0\right)}-\Re\left[\eta'_{\mathrm{v}}\right]\right)>0$. Consequently, inspection of (\ref{eq:COMP:equiv}) reveals

\begin{equation*}
\Re\left[\eta'_{\mathrm{v}}\right]=\eta^{\left(0\right)}_{\mathrm{v}},
\end{equation*}
as required.
\end{proof}

An interesting consequence of this result is that, setting $\tau=\tau_{i}$ and then $i=0$ in (\ref{eq:COMP:imtand}), using (\ref{eq:FORM:finalderiv}), (\ref{eq:COMP:toprove}) and (\ref{eq:COMP:retanbot}), it is evident that

\begin{equation}
\tanh\left(\Im\left[\eta'_{\mathrm{v}}\right]\right)=\frac{\partial\eta_{\mathrm{v}}}{\partial\tau}\left(\tau=\tau_{0}\right),
\end{equation}
so that we may write

\begin{equation}
\eta'_{\mathrm{v}}=\eta^{\left(0\right)}_{\mathrm{v}}+\rmi\tanh^{-1}\left[\frac{\partial\eta_{\mathrm{v}}}{\partial\tau}\left(\tau=\tau_{0}\right)\right].
\end{equation}
Hence, the imaginary part of the complex-valued approximation to the
scattering phase shift obtained in the complex Kohn method can be used as a
measure of the susceptibility of the corresponding generalized Kohn
calculation to Schwartz-type behaviour. In the case of a calculation involving
the exact scattering wavefunction, the imaginary part of \psc would be zero
and the corresponding generalized Kohn calculation of \ps would be independent
of $\tau$.

\section{The relationship to previous work on the generalized Kohn method}
\label{sec:familiar}

In this section, we briefly relate our current work to earlier studies of
Kohn-type methods and Schwartz singularities
\cite{CooperArmourPlummer2009,Schwartz1961,Schwartz1961b,Nesbet1968,Brownstein1968,Shimamura1971,Nesbet1978,Takatsuka1979}.
This is intended as a guide and we do not attempt the full rigorous approach of the
rest of the article. For clarity, we use notation similar to that of Burke and
Joachain \cite{BurkeJoachain1995}. The multichannel extension of this notation
is given, for example, by Nesbet \cite{Nesbet1980} and Lucchese
\cite{Lucchese1989}. Note that in \cite{BurkeJoachain1995} the
operator, $2\left(E-\hat{H}\right)$, for a short-range radial potential scattering
problem is considered rather than $\left(\hat{H}-E\right)$ as used here. For completeness, we note that alternative versions of the Kohn method have been developed in terms of a Feshbach projection operator formalism \cite{Feshbach1962} and have been found \cite{Chung1971} to give anomaly-free results. More recently these methods have been revived and further developed \cite{Bhatia1993} and have produced accurate phase shifts for low energy electron hydrogen atom scattering \cite{Bhatia2001}.

The traditional approach is to separate the `closed' part of the matrix, $A$,
from the `open' part which comprises the first row and column and involves the functions, $\bar{S}$ and
$\bar{C}$. This may be done by inverting the `closed' matrix ($A^{\left(1\right)}_{\left(1\right)}$ in \ref{app:theta}) or, equivalently, by diagonalizing it. This separation aids the analysis but may not necessarily be
carried out in practical calculations as each term in $A$ is
energy-dependent, although the inverse or diagonalization need only be
calculated once for all $\tau$. The closed terms are then `folded' into the
open channel matrix elements as optical potentials \cite{Burke1977,BurkeJoachain1995,Nesbet1980} so that the generalized Kohn problem is reduced to one involving
matrices of a dimension equal to the number of open channels. If the eigenvalues of 
$A^{\left(1\right)}_{\left(1\right)}$ are taken as $\left(E_f-E\right)$, $f=1,\dots,M$,
then we define

\numparts
\begin{eqnarray}\label{eq:fam:defL}
\bar{L}_{11} &=&\langle\bar{S}\psi_{\mathrm{G}},\bar{S}\psi_{\mathrm{G}}\rangle
- \sum_{f=1}^M \frac {\langle\bar{S}\psi_{\mathrm{G}},\chi^{\mathrm{D}}_{f}\rangle
  \langle\chi^{\mathrm{D}}_{f},\bar{S}\psi_{\mathrm{G}}\rangle}
{(E_f - E)},\\
\bar{L}_{22} &=& \langle\bar{C}\psi_{\mathrm{G}},\bar{C}\psi_{\mathrm{G}}\rangle
- \sum_{f=1}^M \frac{\langle\bar{C}\psi_{\mathrm{G}},\chi^{\mathrm{D}}_{f}\rangle
  \langle\chi^{\mathrm{D}}_{f},\bar{C}\psi_{\mathrm{G}}\rangle} 
{(E_f - E)},\\
\bar{L}_{12} &=& \langle\bar{S}\psi_{\mathrm{G}},\bar{C}\psi_{\mathrm{G}}\rangle
- \sum_{f=1}^M \frac {\langle\bar{S}\psi_{\mathrm{G}},\chi^{\mathrm{D}}_{f}\rangle
  \langle\chi^{\mathrm{D}}_{f},\bar{C}\psi_{\mathrm{G}}\rangle} 
{(E_f - E)},\\
\bar{L}_{21} &=& \langle\bar{C}\psi_{\mathrm{G}},\bar{S}\psi_{\mathrm{G}}\rangle
- \sum_{f=1}^M \frac{\langle\bar{C}\psi_{\mathrm{G}},\chi^{\mathrm{D}}_{f}\rangle
  \langle\chi^{\mathrm{D}}_{f},\bar{S}\psi_{\mathrm{G}}\rangle}
{(E_f - E)},
\end{eqnarray}
\endnumparts
where the $\chi^{\mathrm{D}}_{f}$ are the diagonalized linear combinations of the
$\chi_{i}$.

In an obvious notation, the $\bar{L}_{ij}$ are related to their $\tau=0$ values by

\begin{eqnarray}\label{eq:detl}\fl
\left( \begin{array} {cc}
\bar{L}_{11} & \bar{L}_{12} \\ \bar{L}_{21} & \bar{L}_{22} \end{array} \right)
= \left( \begin{array} {cc}
\cos\left(\tau\right) & \sin\left(\tau\right) \\ - \sin\left(\tau\right) & \cos\left(\tau\right) \end{array} \right)
\left( \begin{array} {cc}
{L}_{11} & {L}_{12} \\ {L}_{21} & {L}_{22} \end{array} \right)
\left( \begin{array} {cc}
\cos\left(\tau\right) &  - \sin\left(\tau\right) \\ \sin\left(\tau\right) & \cos\left(\tau\right) \end{array} \right)
\end{eqnarray}
and the generalized Kohn result is

\begin{equation}\label{eq:tandelta}
\tan\left(\eta_{\mathrm{v}}-\tau+c\right)= - \frac{\bar{L}_{21}}{\bar{L}_{22}}
- \frac{1}{\tilde{k} \bar{L}_{22}} \det\left(L\right),
\end{equation}
in which

\begin{equation}
\det\left(L\right) = \bar{L}_{11}\bar{L}_{22} - \bar{L}_{12}\bar{L}_{21}
\end{equation}
is independent of $\tau$ from (\ref{eq:detl}). We note that $\left(\bar{L}_{12}-
\bar{L}_{21}\right)$ and $\left(\bar{L}_{22}+\bar{L}_{11}\right)$ are also independent of $\tau$.

The problem with this approach is that it introduces singularities at the
energies, \newline$E=E_f$, which need to be accounted for. In
practical calculations, $E$ will certainly range across one or more of these
poles as the variational open channel functions are required to take into account only the
asymptotic behaviour of the exact open channel functions (although more
sophisticated functions than this may be chosen). When the Kohn method was first introduced
these poles were briefly considered as causes of the Schwartz-type anomalies.
However, Nesbet \cite{Nesbet1980,Nesbet1968} showed that (\ref{eq:tandelta}) is nonsingular at these energies; the second order poles in $\det\left(L\right)$ cancel leaving first order poles which cancel with those in the denominator.

The framework of this article avoids the universal introduction of these poles and also
avoids the need for closed channel diagonalization to relate the analysis to
results of practical calculations. Intermediate poles are limited to
expressions that use (\ref{eq:FORM:gsim}) to determine $\Gamma\left(k_{\mathrm{s}}\right)$,
and we have described a nonsingular expression for
$\Gamma\left(k_{\mathrm{s}}\right)$ at the end of subsection \ref{ss:anom}, using
the work of \ref{app:theta}.

We may illustrate the relationship between the current approach and the traditional approach by
considering an expansion of $\det(A)$ with the closed matrix,
$A^{\left(1\right)}_{\left(1\right)}$, replaced by the diagonal matrix of
eigenvalues, viz.

\begin{eqnarray}\label{eq:traddet}\nonumber
\det \left(A\right) &=& \langle\bar{C}\psi_{\mathrm{G}},\bar{C}\psi_{\mathrm{G}}\rangle
\prod_{f = 1}^M \left(E_f - E\right)\\
&-& \sum_{f = 1}^M
\langle\bar{C}\psi_{\mathrm{G}},\chi^{\mathrm{D}}_{f}\rangle
  \langle\chi^{\mathrm{D}}_{f},\bar{C}\psi_{\mathrm{G}}\rangle
\prod_{f' = 1, f' \neq f}^M (E_{f'} - E) 
\end{eqnarray}
or, introducing the intermediate poles,
\begin{equation}\label{eq:traddet2}
\det \left(A\right) = \left[\prod_{f = 1}^M \left(E_f - E\right)\right] \bar{L}_{22}.
\end{equation}

Forming a similar expression for the matrix, $\tilde{A}$, and writing 

\begin{equation}\label{eq:eigprod}
F\left(k\right) = \prod_{f = 1}^M \left(E_f - E\right) = \det\left(A^{\left(1\right)}_{\left(1\right)}\right),
\end{equation}
we have the correspondence,

\numparts
\begin{eqnarray}\label{eq:relate}
\mathcal{A}\left(k\right) &\longleftrightarrow& F\left(k\right)L_{11},\\
\mathcal{C}\left(k\right) &\longleftrightarrow& F\left(k\right)L_{22},\\
\mathcal{B}\left(k\right) &\longleftrightarrow& -F\left(k\right)\left(L_{12} + L_{21}\right),\\
\tilde{\mathcal{A}}\left(k\right) &\longleftrightarrow& F\left(k\right)L_{12}.
\end{eqnarray}
\endnumparts

Also,

\begin{equation}
2\tilde{\mathcal{A}}\left(k\right) + \mathcal{B}\left(k\right) \longleftrightarrow F\left(k\right)(L_{12} - L_{21}) = F\left(k\right)\tilde{k},
\end{equation}
so that each instance of $k=k_{\mathrm{s}}$ corresponds to $E=E_f$ for some $f$, and 

\begin{equation}
\Gamma\left(k\right) \longleftrightarrow F\left(k\right)\frac{\det\left(L\right)}{\tilde{k}}.
\end{equation} 

We have reproduced all of the results for the generalized and complex Kohn methods
derived in the preceding sections independently using the $L_{ij}$ formalism, though not always with the same strict rigour. We note that the relatively laborious proofs in Appendices A and B are required for strict avoidance of intermediate poles, otherwise the algebra is of equivalent complexity. We note that in the case of the more familiar $T$-matrix and $S$-matrix versions of the complex Kohn method, with open channel functions
\begin{equation}
\bar{S} + a^{\mathrm{T}}_\mathrm{t} (\bar{C} + \rmi\bar{S})
\end{equation}
and
\begin{equation}
(\bar{C} - \rmi\bar{S}) - a^\mathrm{S}_\mathrm{t} (\bar{C} + \rmi\bar{S}),
\end{equation}
respectively, the behaviour with respect to $\tau$ is the same as presented
here, with correct to second order variational estimates

\begin{equation}
a^\mathrm{S}_\mathrm{v} = 1 + 2\rmi a^\mathrm{T}_\mathrm{v}
\end{equation}
and 

\begin{equation}
a^\mathrm{T}_\mathrm{v} = \rmi (a'_\mathrm{v})^\ast 
\end{equation}
as expected. Here, $a^\mathrm{T}_\mathrm{t}$ and $a^\mathrm{S}_\mathrm{t}$
play a role analogous to $a'_{\mathrm{t}}$ in
(\ref{eq:COMP:complextrialwave}), and we have defined 
\begin{equation}
a'_{\mathrm{v}} = a'_{\mathrm{t}} + \frac{\rmi}{\tilde{k}}\mathcal{I}'\left[\breve{\Psi}_{\mathrm{t}}\right].
\end{equation}
This leads to the $T$-matrix method variational estimate of the phase shift, $\eta^\mathrm{T}_\mathrm{v}$, having the form 
\begin{equation}
\eta^\mathrm{T}_{\mathrm{v}}=\eta^{\left(0\right)}_{\mathrm{v}}-\rmi\tanh^{-1}\left[\frac{\partial\eta_{\mathrm{v}}}{\partial\tau}\left(\tau=\tau_{0}\right)\right].
\end{equation}

We end this section with a few remarks on the earlier ways the generalized Kohn
method was used to avoid anomalous behaviour
\cite{Nesbet1968,Shimamura1971,Nesbet1978,Takatsuka1979}, as summarized by
Nesbet \cite{Nesbet1980}. These various methods certainly avoid
the anomalous singularities but they generally do so by attempting to maximize
the absolute value of $\det\left(A\right)$ or $\bar{L}_{22}$ as a function of
$\tau$. We may take the derivative of (\ref{eq:COMP:determinant}) with respect to $\tau$ and set it to zero.
Denoting by $\tau_{\mathrm{d}}$ any value making \deter stationary with respect to $\tau$, we then obtain the expression

\begin{equation}\label{eq:tandeta}
\left[\mathcal{A}\left(k\right)-\mathcal{C}\left(k\right)\right]\sin\left(2\tau_{\mathrm{d}}\right)+\mathcal{B}\left(k\right)\cos\left(2\tau_{\mathrm{d}}\right)=0.
\end{equation}
Thus, if $k\neq k_{\mathrm{h}}$, there are exactly two values of $\tau\in\left[0,\pi\right)$ making \deter stationary, separated by $\pi/2$. We label these values $\tau_{\mathrm{d}_1}$ and $\tau_{\mathrm{d}_2}$, and note that they are distinct from $\tau_0$ and $\tau_1$. We abbreviate by $\eta^{\left(\mathrm{d}_1\right)}_{\mathrm{v}}$ and $\eta^{\left(\mathrm{d}_2\right)}_{\mathrm{v}}$ the values of $\eta_{\mathrm{v}}\left(\tau_{\mathrm{d}_1}\right)$ and $\eta_{\mathrm{v}}\left(\tau_{\mathrm{d}_2}\right)$ respectively. In general, including the particular case where the two stationary values of $\det\left(A\right)$ are of equal magnitude and opposite sign, $\eta^{\left(\mathrm{d}_1\right)}_{\mathrm{v}}$ and $\eta^{\left(\mathrm{d}_2\right)}_{\mathrm{v}}$ are not equal (see \cite{Cooper:thesis}, which contains material supplementary to that presented here and in \cite{CooperArmourPlummer2009}). Certain of the anomaly-free methods have developed additional techniques to cope with this possibility \cite{Nesbet1980,Nesbet1978}.
  
Remembering that $\det\left(A\right)$ is periodic in $\tau$, the conditions for stationary values with respect to $\tau$
are related to the conditions for existence of the Kohn singularities. The two stationary values of $\det \left( A\right)$ will, in general, comprise one maximum and one minimum. These extrema will either have the same sign (no singularities), or exactly one of them will be zero (exactly one singularity), or they will have different signs (exactly two singularities). This is obvious qualitatively, and a detailed algebraic description of the behaviour of $\det\left(A\right)$ with respect to $\tau$ is available from the authors. From the argument we have proposed \cite{CooperArmourPlummer2009} regarding the existence of anomaly-free singularities, in the case above where exactly one singularity exists, for a sufficiently accurate trial wavefunction we would expect it to correspond to an anomaly-free calculation of the phase shift approximation. In the case above where exactly two singularities exist, for a sufficiently accurate trial wavefunction we would expect one of these singularities to be anomaly-free and the other to be anomalous, in the sense discussed earlier \cite{CooperArmourPlummer2009}.

Some of the above referenced anomaly-free methods attempt to use $\tau$
as a variational parameter once the anomalous region has been avoided. This is done either 
to help increase the phase shift, though we note that the Kohn methods provide 
stationary but not bounded variational estimates, or otherwise to try to 
improve the `quality' of the trial wavefunction according to various criteria 
put forward. This is in contrast to the current work, in which we argue that
$\tau$ should ideally be an arbitrary parameter and we look for the minimum
value of \fderiv. The `quality' of the scattering wavefunction then depends on the forms of the $S$, $C$ and $\{\chi_{i}\}$ functions when the projectile is close to the target. We suggest that,
in addition to the direct relationship with the complex Kohn method derived above,
improvements in computing power and computational science over the last few
decades justify this approach, with nonlinear parameters in the bulk trial
functions available for use as additional variational parameters.

Cooper \cite{Cooper:thesis} has carried out some studies of the behaviour of \fderiv at $\tau=\tau_{0}$, as a function of $k$. This behaviour is related to the separation of the anomaly-free and anomalous singularities as a function of $k$ \cite{CooperArmourPlummer2009,Cooper:thesis}. As defined \cite{Cooper:thesis}, this separation is a value in the range
$\left[0,\pi/2\right]$. When the two singularities are separated by $\pi/2$, their $\tau$ values satisfy the condition (\ref{eq:FORM:xyeq}) required of $\tau_0$ and $\tau_1$. However,
\fderiv at $\tau=\tau_{0}$ is generally smaller as a function of $k$ when the separation of the two singularities is also small \cite{Cooper:thesis}. 

We note that the earlier anomaly-free methods have been extended to the multichannel case \cite{Nesbet1980,Nesbet1978,Lucchese1989,Takatsuka2}. The complex Kohn method is being applied to various calculations of
electron polyatomic molecule scattering (see, for example, \cite{Rescigno2006,Rescigno2007}). We are currently extending the framework and analysis of this article to the multichannel case and we aim to develop the analysis of the anomaly-free singularities as part of the multichannel work. 

\section{Concluding remarks}\label{s:conclusions}

We have shown that, in the case of phase shift calculations for low energy \ehmol scattering, the complex Kohn method is equivalent to a particular optimization of the generalized Kohn method. Further, we have established a number of interesting results regarding the appearance of anomalous behaviour in our generalized Kohn calculations. Specifically, we have found that anomalies which appear when only $\tau$ is varied can be explained from purely analytic considerations; they are intrinsic to the Kohn method itself and do not, as has previously been suggested \cite{ArmourHumberston1991}, arise from matrix ill-conditioning or errors due to limited computational precision. Our analysis describes analytically the behaviour of the phase shift over the {\it entire} range of $\tau$. This makes it possible to give a full description, for the first time, of any anomalous behaviour that results from the variation of $\tau$.

By obtaining an analytic expression for \fderiv, we have explained the mathematical origin of the anomaly-free singularities identified in our earlier article \cite{CooperArmourPlummer2009}. This result complements the physical argument for the existence of these singularities given in that article. 

We have demonstrated that there is a particular class of anomalies that are necessarily avoided in our complex Kohn calculations. However, it is important to note that this method, as we have implemented it here, avoids only those
anomalies encountered in generalized Kohn calculations by varying $\tau$ and keeping other free parameters, such as $k$, $R$ and the nonlinear parameters in the trial function,
fixed. Anomalous results characterized by unusually large values of \kderiv,
say, could still arise even in the complex Kohn method. We have not developed an
explicit expression for this derivative in the way that we have here for \fderiv and
\cderiv; strictly speaking, we have not even considered in detail whether \kderiv exists in the sense of demonstrating that $\Re\left[\eta'_{\mathrm{v}}\right]$ is a differentiable function of $k$. However, were an analytic expression for \kderiv available, it is reasonable to conclude that it would explain the persistent anomalous behaviour discussed in \cite{CooperArmourPlummer2009}.

\ack
JNC wishes to thank John Humberston and Chris Franklin for valuable discussions. This work is supported by EPSRC (UK) grant EP/C548019/1.

\appendix

\section{The $\Theta$ function}\label{app:theta}

For nonsingular $A$, we claim

\begin{equation}
\det\left(A\right)\mathcal{I}\left[\Psi_{\mathrm{t}}\right]=\Theta\left(k,\cancel{\tau}\right),
\end{equation}
where we have used a self-evident notation to denote that $\Theta\left(k,\cancel{\tau}\right)$ is independent of $\tau$. 

\begin{proof}
In the following argument, we will write $\Theta=\Theta\left(k,\tau\right)$ before explicitly proving the independence of $\Theta$ from $\tau$. Henceforth, it will be convenient to consider only the case $M\geq3$ in (\ref{eq:INT:trialwaveEXPL}). It is straightforward to show that the result (\ref{eq:GEN:theta}) is satisfied for\linebreak $M<3$ by explicitly inverting the matrix, $A$, allowing $\mathcal{I}\left[\Psi_{\mathrm{t}}\right]$ to be found. Throughout, we will implicitly make use of the Hermiticity properties (\ref{eq:INT:HERM2})-(\ref{eq:INT:HERM5}).

We denote by $\tilde{A}_{\left(j\right)}$ the $\left(M+1 \times M+1\right)$ matrix formed by replacing the $j^{\mathrm{th}}$ column of $A$ (\ref{eq:matA}) by $-b$ (\ref{eq:vecB}). We will denote by $A^{\left(i\right)}_{\left(j\right)}$ the $\left(M \times M\right)$ matrix formed by removing the $i^{\mathrm{th}}$ row and $j^{\mathrm{th}}$ column of $A$. The row and column indices of $A$ range from $1$ to $M+1$. By assumption, $A$ is nonsingular, so that using (\ref{eq:APT:redI}) and Cramer's rule, which states that

\begin{equation}\label{eq:APT:cram1}
 a_{\mathrm{t}}=\frac{\det\left(\tilde{A}_{\left(1\right)}\right)}{\det\left(A\right)}
\end{equation}
and, for $1<j\leq M$,

\begin{equation}
 p_{j}=\frac{\det\left(\tilde{A}_{\left(j+1\right)}\right)}{\det\left(A\right)},
\end{equation}
the product, $\det\left(A\right)\mathcal{I}\left[\Psi_{\mathrm{t}}\right]$, can be written

\begin{eqnarray}\fl
\nonumber\Theta\left(k,\tau\right)&=&\det\left(A\right)\mathcal{I}\left[\Psi_{\mathrm{t}}\right]=\det\left(A\right)\langle\bar{S}\psi_{\mathrm{G}},\bar{S}\psi_{\mathrm{G}}\rangle+\det\left(\tilde{A}_{\left(1\right)}\right)\langle\bar{S}\psi_{\mathrm{G}},\bar{C}\psi_{\mathrm{G}}\rangle\\
\label{eq:APT:A1}\fl&+&\sum_{j=1}^{M} \det\left(\tilde{A}_{\left(j+1\right)}\right)\langle\bar{S}\psi_{\mathrm{G}},\chi_{j}\rangle.
\end{eqnarray}
The Laplace expansion of \deter along column $1$ of $A$ is

\begin{equation}\label{eq:APT:A2}\fl
\det\left(A\right)=\langle\bar{C}\psi_{\mathrm{G}},\bar{C}\psi_{\mathrm{G}}\rangle\det\left(A^{\left(1\right)}_{\left(1\right)}\right)+\sum_{i=1}^{M}\left(-1\right)^{i}\langle\bar{C}\psi_{\mathrm{G}},\chi_{i}\rangle\det\left(A^{\left(i+1\right)}_{\left(1\right)}\right),
\end{equation}
while the expansion of $\det\left(\tilde{A}_{\left(j\right)}\right)$ along column $j$ of $\tilde{A}_{\left(j\right)}$ is 

\begin{equation}\label{eq:APT:A3}\fl
\det\left(\tilde{A}_{\left(j\right)}\right)=\left(-1\right)^j\left[\langle\bar{C}\psi_{\mathrm{G}},\bar{S}\psi_{\mathrm{G}}\rangle\det\left(A^{\left(1\right)}_{\left(j\right)}\right)+\sum_{i=1}^{M}\left(-1\right)^{i}\langle\bar{S}\psi_{\mathrm{G}},\chi_{i}\rangle\det\left(A^{\left(i+1\right)}_{\left(j\right)}\right)\right].
\end{equation}
Now, using (\ref{eq:tautrans}), we obtain

\begin{eqnarray}\nonumber\
&&\langle\bar{S}\psi_{\mathrm{G}},\bar{S}\psi_{\mathrm{G}}\rangle\langle\bar{C}\psi_{\mathrm{G}},\bar{C}\psi_{\mathrm{G}}\rangle-\langle\bar{S}\psi_{\mathrm{G}},\bar{C}\psi_{\mathrm{G}}\rangle\langle\bar{C}\psi_{\mathrm{G}},\bar{S}\psi_{\mathrm{G}}\rangle\\
\label{eq:APT:A4}=&&\langle S\psi_{\mathrm{G}}, S\psi_{\mathrm{G}}\rangle\langle C\psi_{\mathrm{G}},C\psi_{\mathrm{G}}\rangle-\langle S\psi_{\mathrm{G}},C\psi_{\mathrm{G}}\rangle\langle C\psi_{\mathrm{G}},S\psi_{\mathrm{G}}\rangle,
\end{eqnarray}
which is independent of $\tau$. Hence, by noting that $\det\left(A^{\left(1\right)}_{\left(1\right)}\right)$ is also independent of $\tau$, then defining

\begin{eqnarray}
 \nonumber\Theta_{0}\left(k,\cancel{\tau}\right)&=&\left(\langle S\psi_{\mathrm{G}}, S\psi_{\mathrm{G}}\rangle\langle C\psi_{\mathrm{G}},C\psi_{\mathrm{G}}\rangle-\langle S\psi_{\mathrm{G}},C\psi_{\mathrm{G}}\rangle\langle C\psi_{\mathrm{G}},S\psi_{\mathrm{G}}\rangle\right)\\
\label{eq:APT:A5}&\times&\det\left(A^{\left(1\right)}_{\left(1\right)}\right)
\end{eqnarray}
and combining (\ref{eq:APT:A1}), (\ref{eq:APT:A2}), (\ref{eq:APT:A3}) and (\ref{eq:APT:A4}), we can write

\begin{eqnarray}
\nonumber\fl\Theta\left(k,\tau\right)-\Theta_{0}\left(k,\cancel{\tau}\right)&=&\langle\bar{S}\psi_{\mathrm{G}},\bar{S}\psi_{\mathrm{G}}\rangle\sum_{i=1}^{M}\left(-1\right)^{i}\langle\bar{C}\psi_{\mathrm{G}},\chi_{i}\rangle\det\left(A^{\left(i+1\right)}_{\left(1\right)}\right)\\
\nonumber\fl&-&\langle\bar{S}\psi_{\mathrm{G}},\bar{C}\psi_{\mathrm{G}}\rangle\sum_{i=1}^{M}\left(-1\right)^{i}\langle\bar{S}\psi_{\mathrm{G}},\chi_{i}\rangle\det\left(A^{\left(i+1\right)}_{\left(1\right)}\right)\\
\nonumber\fl&+&\sum_{j=1}^{M}\langle\bar{S}\psi_{\mathrm{G}},\chi_{j}\rangle\left(-1\right)^{j+1} \left[\langle\bar{C}\psi_{\mathrm{G}},\bar{S}\psi_{\mathrm{G}}\rangle\det\left(A^{\left(1\right)}_{\left(j+1\right)}\right)\vphantom{\sum_{i=1}^{M}}\right.\\
\label{eq:APT:A6}\fl&+&\left.\sum_{i=1}^{M}\left(-1\right)^{i}\langle\bar{S}\psi_{\mathrm{G}},\chi_{i}\rangle\det\left(A^{\left(i+1\right)}_{\left(j+1\right)}\right)\vphantom{\sum_{i=1}^{M}}\right].
\end{eqnarray}
It then remains to be shown that the right hand side of (\ref{eq:APT:A6}) is independent of $\tau$. 

Using the fact that $\det\left(M\right)=\det\left(M^{\top}\right)$ for any square matrix, $M$, together with the fact that $A$ is symmetric, we deduce that $\det\left(A^{\left(p\right)}_{\left(q\right)}\right)=\det\left(A^{\left(q\right)}_{\left(p\right)}\right)$. We can then rewrite (\ref{eq:APT:A6}) as 

\begin{eqnarray}
\nonumber\fl\Theta\left(k,\tau\right)-\Theta_{0}\left(k,\cancel{\tau}\right)&=&\left[\sum_{i=1}^{M}\left(-1\right)^{i}\det\left(A^{\left(i+1\right)}_{\left(1\right)}\right)\vphantom{\sum_{i=1}^M}\right.\\
\nonumber\fl&\times&(\langle\bar{S}\psi_{\mathrm{G}},\bar{S}\psi_{\mathrm{G}}\rangle\langle\bar{C}\psi_{\mathrm{G}},\chi_{i}\rangle\\
\nonumber\fl&-&\langle\bar{S}\psi_{\mathrm{G}},\bar{C}\psi_{\mathrm{G}}\rangle\langle\bar{S}\psi_{\mathrm{G}},\chi_{i}\rangle\\
\nonumber\fl&-&\left.\langle\bar{C}\psi_{\mathrm{G}},\bar{S}\psi_{\mathrm{G}}\rangle\langle\bar{S}\psi_{\mathrm{G}},\chi_{i}\rangle)\vphantom{\sum_{i=1}^M}\right]\\\label{eq:APT:A7}
\fl&+&\sum_{i=1}^M\sum_{j=1}^M\left(-1\right)^{i+j+1}\det\left(A^{\left(i+1\right)}_{\left(j+1\right)}\right)\langle\bar{S}\psi_{\mathrm{G}},\chi_{i}\rangle\langle\bar{S}\psi_{\mathrm{G}},\chi_{j}\rangle.
\end{eqnarray}
We next define the $\left(M\times M\right)$ matrix, $X=A^{\left(1\right)}_{\left(1\right)}$. Further, we denote by $X^{\left(i\right)}_{\left(j\right)}$ the $\left(M-1 \times M-1\right)$ matrix formed by removing the $i^{\mathrm{th}}$ row and $j^{\mathrm{th}}$ column of $X$. Further, for $i\neq p$ and $j\neq q$, we denote by $X^{\left(i,p\right)}_{\left(j,q\right)}$ the $\left(M-2 \times M-2\right)$ matrix formed by removing the $i^{\mathrm{th}}$ and $p^{\mathrm{th}}$ rows and $j^{\mathrm{th}}$ and $q^{\mathrm{th}}$ columns of $X$. The row and column indices of $X$ range from $1$ to $M$. The elements of $X$, $X^{\left(i\right)}_{\left(j\right)}$ and $X^{\left(i,p\right)}_{\left(j,q\right)}$ are independent of $\tau$. If $i=p$ or $j=q$, we define $X^{\left(i,p\right)}_{\left(j,q\right)}$ to be the $\left(M-2 \times M-2\right)$ matrix of zeros. We find that

\begin{equation}\label{eq:APT:A8}
 \det\left(A^{\left(i+1\right)}_{\left(1\right)}\right)=\sum_{j=1}^{M}\left(-1\right)^{j+1} \det\left(X^{\left(i\right)}_{\left(j\right)}\right)\langle\bar{C}\psi_{\mathrm{G}},\chi_{j}\rangle.
\end{equation}
Moreover, after careful consideration we have 

\begin{eqnarray}
 \nonumber\det\left(A^{\left(i+1\right)}_{\left(j+1\right)}\right)&=&\det\left(X^{\left(i\right)}_{\left(j\right)}\right)\langle\bar{C}\psi_{\mathrm{G}},\bar{C}\psi_{\mathrm{G}}\rangle\\
\nonumber&+&\left[\sum_{p=1}^M\sum_{q=1}^M\left(-1\right)^{p+q+1+\sigma_{ip}+\sigma_{jq}}\vphantom{\sum_{p=1}^M}\right.\\
\label{eq:APT:A9}&\times&\left.\vphantom{\sum_{p=1}^M}\det\left(X^{\left(i,p\right)}_{\left(j,q\right)}\right)\langle\bar{C}\psi_{\mathrm{G}},\chi_{p}\rangle\langle\bar{C}\psi_{\mathrm{G}},\chi_{q}\rangle\right],
\end{eqnarray}
where terms of the form $\sigma_{ab}$ have the definition

\begin{eqnarray}\label{eq:APT:sigma}
    \sigma_{ab} = \left\{\begin{array}{ll}
                     0    & \quad (a\geq b) \\
                     1    & \quad (a < b) \\
                     \end{array} \right. .
\end{eqnarray}
Using (\ref{eq:APT:A8}) and (\ref{eq:APT:A9}), we rewrite (\ref{eq:APT:A7}) as

\begin{equation}\label{eq:APT:A10}
 \Theta\left(k,\tau\right)=\Theta_{0}\left(k,\cancel{\tau}\right)+\Theta_{1}\left(k,\tau\right)+\Theta_{2}\left(k,\tau\right),
\end{equation}
where

\begin{eqnarray}
\nonumber\fl\Theta_{1}\left(k,\tau\right)=\sum_{i=1}^M\sum_{j=1}^M\left(-1\right)^{i+j+1}\det\left(X^{\left(i\right)}_{\left(j\right)}\right)&\times&\left(\vphantom{\sum_{i=1}^M}\langle\bar{S}\psi_{\mathrm{G}},\bar{S}\psi_{\mathrm{G}}\rangle\langle\bar{C}\psi_{\mathrm{G}},\chi_{i}\rangle\langle\bar{C}\psi_{\mathrm{G}},\chi_{j}\rangle\right.\\
\nonumber\fl&-&\langle\bar{S}\psi_{\mathrm{G}},\bar{C}\psi_{\mathrm{G}}\rangle\langle\bar{S}\psi_{\mathrm{G}},\chi_{i}\rangle\langle\bar{C}\psi_{\mathrm{G}},\chi_{j}\rangle\\
\nonumber\fl&-&\langle\bar{C}\psi_{\mathrm{G}},\bar{S}\psi_{\mathrm{G}}\rangle\langle\bar{S}\psi_{\mathrm{G}},\chi_{i}\rangle\langle\bar{C}\psi_{\mathrm{G}},\chi_{j}\rangle\\
\label{eq:APT:A11}\fl&+&\left.\langle\bar{C}\psi_{\mathrm{G}},\bar{C}\psi_{\mathrm{G}}\rangle\langle\bar{S}\psi_{\mathrm{G}},\chi_{i}\rangle\langle\bar{S}\psi_{\mathrm{G}},\chi_{j}\rangle\vphantom{\sum_{i=1}^M}\right)
\end{eqnarray}
and

\begin{eqnarray}
\nonumber\fl\Theta_{2}\left(k,\tau\right)=\sum_{i=1}^M\sum_{j=1}^M\sum_{p=1}^M\sum_{q=1}^M&&\left[\vphantom{\sum_{i=1}^M}\left(-1\right)^{i+j+p+q+\sigma_{ip}+\sigma_{jq}}\det\left(X^{\left(i,p\right)}_{\left(j,q\right)}\right)\right.\\
\label{eq:APT:A12}\fl&\times&\left.\langle\bar{S}\psi_{\mathrm{G}},\chi_{i}\rangle \langle\bar{S}\psi_{\mathrm{G}},\chi_{j}\rangle \langle\bar{C}\psi_{\mathrm{G}},\chi_{p}\rangle \langle\bar{C}\psi_{\mathrm{G}},\chi_{q}\rangle\vphantom{\sum_{i=1}^M}\right].
\end{eqnarray}
We will now show that $\Theta_{1}\left(k,\tau\right)$ and $\Theta_{2}\left(k,\tau\right)$ are each independent of $\tau$. 

Considering (\ref{eq:APT:A11}), since $X$ is symmetric, we note that $\det\left(X^{\left(i\right)}_{\left(j\right)}\right)=\det\left(X^{\left(j\right)}_{\left(i\right)}\right)$. Using (\ref{eq:tautrans}), when the summation over $i$ and $j$ in (\ref{eq:APT:A11}) is carried out, a number of terms cancel. We then find 

\begin{eqnarray}
\nonumber\fl\Theta_{1}\left(k,\cancel{\tau}\right)=\sum_{i=1}^M\sum_{j=1}^M\left(-1\right)^{i+j+1}\det\left(X^{\left(i\right)}_{\left(j\right)}\right)&\times&\left(\vphantom{\sum_{i=1}^M}\langle S\psi_{\mathrm{G}}, S\psi_{\mathrm{G}}\rangle\langle C\psi_{\mathrm{G}},\chi_{i}\rangle\langle C\psi_{\mathrm{G}},\chi_{j}\rangle\right.\\
\nonumber\fl&-&\langle S\psi_{\mathrm{G}}, C\psi_{\mathrm{G}}\rangle\langle S\psi_{\mathrm{G}},\chi_{i}\rangle\langle C\psi_{\mathrm{G}},\chi_{j}\rangle\\
\nonumber\fl&-&\langle C\psi_{\mathrm{G}}, S\psi_{\mathrm{G}}\rangle\langle S\psi_{\mathrm{G}},\chi_{i}\rangle\langle C\psi_{\mathrm{G}},\chi_{j}\rangle\\
\label{eq:APT:A13}\fl&+&\left.\langle C\psi_{\mathrm{G}}, C\psi_{\mathrm{G}}\rangle\langle S\psi_{\mathrm{G}},\chi_{i}\rangle\langle S\psi_{\mathrm{G}},\chi_{j}\rangle\vphantom{\sum_{i=1}^M}\right),
\end{eqnarray}
which is independent of $\tau$. Applying the same method to (\ref{eq:APT:A12}), with a little work we can write

\begin{eqnarray}
\nonumber\fl\Theta_{2}\left(k,\cancel{\tau}\right)=\sum_{i=1}^M\sum_{j=1}^M\sum_{p=1}^M\sum_{q=1}^M&&\left[\vphantom{\sum_{i=1}^M}\left(-1\right)^{i+j+p+q+\sigma_{ip}+\sigma_{jq}}\det\left(X^{\left(i,p\right)}_{\left(j,q\right)}\right)\right.\\
\label{eq:APT:A14}\fl&\times&\left.\langle S\psi_{\mathrm{G}},\chi_{i}\rangle \langle S\psi_{\mathrm{G}},\chi_{j}\rangle \langle C\psi_{\mathrm{G}},\chi_{p}\rangle \langle C\psi_{\mathrm{G}},\chi_{q}\rangle\vphantom{\sum_{i=1}^M}\right],
\end{eqnarray}
which is independent of $\tau$. The cancellation in the summation (\ref{eq:APT:A12}) arises from the fact that

\begin{equation}\label{eq:APT:flip}
\left(-1\right)^{\sigma_{ab}}+\left(-1\right)^{\sigma_{ba}}=0\quad\quad\left(a\neq b\right).
\end{equation}
Finally, combining (\ref{eq:APT:A10}), (\ref{eq:APT:A13}) and (\ref{eq:APT:A14}), we have

\begin{equation}\label{eq:APT:finalresult}
\Theta\left(k,\cancel{\tau}\right)=\Theta_{0}\left(k,\cancel{\tau}\right)+\Theta_{1}\left(k,\cancel{\tau}\right)+\Theta_{2}\left(k,\cancel{\tau}\right),
\end{equation}
so that $\Theta=\Theta\left(k,\cancel{\tau}\right)$, as required.
\end{proof}

In the case of the complex Kohn method, a result similar to (\ref{eq:GEN:theta}) can be derived by a method analogous to that given above. For nonsingular $A'$, we claim

\begin{equation}
\det\left(A'\right)\mathcal{I}'\left[\breve{\Psi}_{\mathrm{t}}\right]=-\Theta\left(k,\cancel{\tau}\right).
\end{equation}

\begin{proof}
We define a function, $\Lambda\left(k,\tau\right)$, such that

\begin{equation}
 \Lambda\left(k,\tau\right) = \det\left(A'\right)\mathcal{I}'\left[\breve{\Psi}_{\mathrm{t}}\right].
\end{equation}
Next, after some manipulation, it is straightforward to show that

\begin{eqnarray}
\nonumber&&\langle\bar{S}\psi_{\mathrm{G}},\bar{S}\psi_{\mathrm{G}}\rangle\langle\bar{T}^{*}\psi_{\mathrm{G}},\bar{T}\psi_{\mathrm{G}}\rangle-\langle\bar{S}\psi_{\mathrm{G}},\bar{T}\psi_{\mathrm{G}}\rangle\langle\bar{T}^{*}\psi_{\mathrm{G}},\bar{S}\psi_{\mathrm{G}}\rangle\\
\nonumber&=&\langle \bar{S}\psi_{\mathrm{G}}, \bar{C}\psi_{\mathrm{G}}\rangle\langle \bar{C}\psi_{\mathrm{G}},\bar{S}\psi_{\mathrm{G}}\rangle-\langle \bar{S}\psi_{\mathrm{G}},\bar{S}\psi_{\mathrm{G}}\rangle\langle \bar{C}\psi_{\mathrm{G}},\bar{C}\psi_{\mathrm{G}}\rangle\\
\label{eq:APT:AC4}&=&\langle S\psi_{\mathrm{G}}, C\psi_{\mathrm{G}}\rangle\langle C\psi_{\mathrm{G}},S\psi_{\mathrm{G}}\rangle-\langle S\psi_{\mathrm{G}},S\psi_{\mathrm{G}}\rangle\langle C\psi_{\mathrm{G}},C\psi_{\mathrm{G}}\rangle,
\end{eqnarray}
where we have used (\ref{eq:APT:A4}). Proceeding in a manner analogous to that used above and adopting the same notation, using (\ref{eq:APT:A5}) it is clear that we can then immediately write

\begin{equation}\label{eq:APT:AC10}
\Lambda\left(k,\tau\right)=-\Theta_{0}\left(k,\cancel{\tau}\right)+\Lambda_{1}\left(k,\tau\right)+\Lambda_{2}\left(k,\tau\right),
\end{equation}
where

\begin{eqnarray}
\nonumber\fl\Lambda_{1}\left(k,\tau\right)=\sum_{i=1}^M\sum_{j=1}^M\left(-1\right)^{i+j+1}\det\left(X^{\left(i\right)}_{\left(j\right)}\right)&\times&\left(\vphantom{\sum_{i=1}^M}\langle\bar{S}\psi_{\mathrm{G}},\bar{S}\psi_{\mathrm{G}}\rangle\langle\bar{T}^{*}\psi_{\mathrm{G}},\chi_{i}\rangle\langle\bar{T}^{*}\psi_{\mathrm{G}},\chi_{j}\rangle\right.\\
\nonumber\fl&-&\langle\bar{S}\psi_{\mathrm{G}},\bar{T}\psi_{\mathrm{G}}\rangle\langle\bar{S}\psi_{\mathrm{G}},\chi_{i}\rangle\langle\bar{T}^{*}\psi_{\mathrm{G}},\chi_{j}\rangle\\
\nonumber\fl&-&\langle\bar{T}^{*}\psi_{\mathrm{G}},\bar{S}\psi_{\mathrm{G}}\rangle\langle\bar{S}\psi_{\mathrm{G}},\chi_{i}\rangle\langle\bar{T}^{*}\psi_{\mathrm{G}},\chi_{j}\rangle\\
\label{eq:APT:AC11}\fl&+&\left.\langle\bar{T}^{*}\psi_{\mathrm{G}},\bar{T}\psi_{\mathrm{G}}\rangle\langle\bar{S}\psi_{\mathrm{G}},\chi_{i}\rangle\langle\bar{S}\psi_{\mathrm{G}},\chi_{j}\rangle\vphantom{\sum_{i=1}^M}\right)
\end{eqnarray}
and

\begin{eqnarray}
\nonumber\fl\Lambda_{2}\left(k,\tau\right)=\sum_{i=1}^M\sum_{j=1}^M\sum_{p=1}^M\sum_{q=1}^M&&\left[\vphantom{\sum_{i=1}^M}\left(-1\right)^{i+j+p+q+\sigma_{ip}+\sigma_{jq}}\det\left(X^{\left(i,p\right)}_{\left(j,q\right)}\right)\right.\\
\label{eq:APT:AC12}\fl&\times&\left.\langle\bar{S}\psi_{\mathrm{G}},\chi_{i}\rangle \langle\bar{S}\psi_{\mathrm{G}},\chi_{j}\rangle \langle \bar{T}^{*}\psi_{\mathrm{G}},\chi_{p}\rangle \langle\bar{T}^{*}\psi_{\mathrm{G}},\chi_{q}\rangle\vphantom{\sum_{i=1}^M}\right],
\end{eqnarray}
$\Lambda_{1}\left(k,\tau\right)$ and $\Lambda_{2}\left(k,\tau\right)$ being analogous to (\ref{eq:APT:A11}) and (\ref{eq:APT:A12}), respectively.
Considering first $\Lambda_{1}\left(k,\tau\right)$, we have

\begin{eqnarray}
\nonumber\fl&&\langle\bar{S}\psi_{\mathrm{G}},\bar{S}\psi_{\mathrm{G}}\rangle\langle\bar{T}^{*}\psi_{\mathrm{G}},\chi_{i}\rangle\langle\bar{T}^{*}\psi_{\mathrm{G}},\chi_{j}\rangle-\langle\bar{S}\psi_{\mathrm{G}},\bar{T}\psi_{\mathrm{G}}\rangle\langle\bar{S}\psi_{\mathrm{G}},\chi_{i}\rangle\langle\bar{T}^{*}\psi_{\mathrm{G}},\chi_{j}\rangle\\
\nonumber\fl&-&\langle\bar{T}^{*}\psi_{\mathrm{G}},\bar{S}\psi_{\mathrm{G}}\rangle\langle\bar{S}\psi_{\mathrm{G}},\chi_{i}\rangle\langle\bar{T}^{*}\psi_{\mathrm{G}},\chi_{j}\rangle+\langle\bar{T}^{*}\psi_{\mathrm{G}},\bar{T}\psi_{\mathrm{G}}\rangle\langle\bar{S}\psi_{\mathrm{G}},\chi_{i}\rangle\langle\bar{S}\psi_{\mathrm{G}},\chi_{j}\rangle\\
\nonumber\fl&=&\langle\bar{S}\psi_{\mathrm{G}},\bar{C}\psi_{\mathrm{G}}\rangle\langle\bar{S}\psi_{\mathrm{G}},\chi_{i}\rangle\langle\bar{C}\psi_{\mathrm{G}},\chi_{j}\rangle-\langle\bar{S}\psi_{\mathrm{G}},\bar{S}\psi_{\mathrm{G}}\rangle\langle\bar{C}\psi_{\mathrm{G}},\chi_{i}\rangle\langle\bar{C}\psi_{\mathrm{G}},\chi_{j}\rangle\\
\nonumber\fl&-&\langle\bar{C}\psi_{\mathrm{G}},\bar{C}\psi_{\mathrm{G}}\rangle\langle\bar{S}\psi_{\mathrm{G}},\chi_{i}\rangle\langle\bar{S}\psi_{\mathrm{G}},\chi_{j}\rangle+\langle\bar{C}\psi_{\mathrm{G}},\bar{S}\psi_{\mathrm{G}}\rangle\langle\bar{S}\psi_{\mathrm{G}},\chi_{i}\rangle\langle\bar{C}\psi_{\mathrm{G}},\chi_{j}\rangle\\
\label{eq:APT:AC13}\fl&+&\rmi\langle\bar{S}\psi_{\mathrm{G}},\bar{S}\psi_{\mathrm{G}}\rangle\left(\langle\bar{C}\psi_{\mathrm{G}},\chi_{i}\rangle\langle\bar{S}\psi_{\mathrm{G}},\chi_{j}\rangle-\langle\bar{S}\psi_{\mathrm{G}},\chi_{i}\rangle\langle\bar{C}\psi_{\mathrm{G}},\chi_{j}\rangle\right).
\end{eqnarray}
When the summation in (\ref{eq:APT:AC11}) is carried out, it is clear that the final terms in the square brackets in (\ref{eq:APT:AC13}) sum to zero. Using (\ref{eq:APT:A11}) and (\ref{eq:APT:A13}), we then have

\begin{eqnarray}
\nonumber\fl\Lambda_{1}\left(k,\cancel{\tau}\right)=-\sum_{i=1}^M\sum_{j=1}^M\left(-1\right)^{i+j+1}\det\left(X^{\left(i\right)}_{\left(j\right)}\right)&\times&\left(\vphantom{\sum_{i=1}^M}\langle S\psi_{\mathrm{G}}, S\psi_{\mathrm{G}}\rangle\langle C\psi_{\mathrm{G}},\chi_{i}\rangle\langle C\psi_{\mathrm{G}},\chi_{j}\rangle\right.\\
\nonumber\fl&-&\langle S\psi_{\mathrm{G}}, C\psi_{\mathrm{G}}\rangle\langle S\psi_{\mathrm{G}},\chi_{i}\rangle\langle C\psi_{\mathrm{G}},\chi_{j}\rangle\\
\nonumber\fl&-&\langle C\psi_{\mathrm{G}}, S\psi_{\mathrm{G}}\rangle\langle S\psi_{\mathrm{G}},\chi_{i}\rangle\langle C\psi_{\mathrm{G}},\chi_{j}\rangle\\
\label{eq:APT:AC14}\fl&+&\left.\langle C\psi_{\mathrm{G}}, C\psi_{\mathrm{G}}\rangle\langle S\psi_{\mathrm{G}},\chi_{i}\rangle\langle S\psi_{\mathrm{G}},\chi_{j}\rangle\vphantom{\sum_{i=1}^M}\right),
\end{eqnarray}
so that

\begin{equation}
 \Lambda_{1}\left(k,\cancel{\tau}\right)=-\Theta_{1}\left(k,\cancel{\tau}\right).
\end{equation}
Next, we consider $\Lambda_{2}\left(k,\tau\right)$. Clearly,

\begin{eqnarray}
\nonumber&&\langle\bar{S}\psi_{\mathrm{G}},\chi_{i}\rangle \langle\bar{S}\psi_{\mathrm{G}},\chi_{j}\rangle \langle \bar{T}^{*}\psi_{\mathrm{G}},\chi_{p}\rangle \langle\bar{T}^{*}\psi_{\mathrm{G}},\chi_{q}\rangle\\
\nonumber&=&\langle\bar{S}\psi_{\mathrm{G}},\chi_{i}\rangle \langle\bar{S}\psi_{\mathrm{G}},\chi_{j}\rangle \langle\bar{S}\psi_{\mathrm{G}},\chi_{p}\rangle \langle\bar{S}\psi_{\mathrm{G}},\chi_{q}\rangle\\
\nonumber&-&\langle\bar{S}\psi_{\mathrm{G}},\chi_{i}\rangle \langle\bar{S}\psi_{\mathrm{G}},\chi_{j}\rangle \langle\bar{C}\psi_{\mathrm{G}},\chi_{p}\rangle \langle\bar{C}\psi_{\mathrm{G}},\chi_{q}\rangle\\
\nonumber&+&\rmi\langle\bar{S}\psi_{\mathrm{G}},\chi_{i}\rangle \langle\bar{S}\psi_{\mathrm{G}},\chi_{j}\rangle\langle\bar{S}\psi_{\mathrm{G}},\chi_{p}\rangle \langle\bar{C}\psi_{\mathrm{G}},\chi_{q}\rangle\\
\label{eq:APT:AC15}&+&\rmi\langle\bar{S}\psi_{\mathrm{G}},\chi_{i}\rangle \langle\bar{S}\psi_{\mathrm{G}},\chi_{j}\rangle\langle\bar{C}\psi_{\mathrm{G}},\chi_{p}\rangle \langle\bar{S}\psi_{\mathrm{G}},\chi_{q}\rangle.
\end{eqnarray}
Using (\ref{eq:APT:flip}), we see that the first term in the expansion (\ref{eq:APT:AC15}) does not give an overall contribution to the sum (\ref{eq:APT:AC12}). For the same reason, each of the final two terms in (\ref{eq:APT:AC15}) also sums to zero in (\ref{eq:APT:AC12}). Hence, using (\ref{eq:APT:A12}) and (\ref{eq:APT:A14}), we then have

\begin{eqnarray}
\nonumber\fl\Lambda_{2}\left(k,\cancel{\tau}\right)=-\sum_{i=1}^M\sum_{j=1}^M\sum_{p=1}^M\sum_{q=1}^M&&\left[\vphantom{\sum_{i=1}^M}\left(-1\right)^{i+j+p+q+\sigma_{ip}+\sigma_{jq}}\det\left(X^{\left(i,p\right)}_{\left(j,q\right)}\right)\right.\\
\label{eq:APT:AC16}\fl&\times&\left.\langle S\psi_{\mathrm{G}},\chi_{i}\rangle \langle S\psi_{\mathrm{G}},\chi_{j}\rangle \langle C\psi_{\mathrm{G}},\chi_{p}\rangle \langle C\psi_{\mathrm{G}},\chi_{q}\rangle\vphantom{\sum_{i=1}^M}\right],
\end{eqnarray}
so that

\begin{equation}
 \Lambda_{2}\left(k,\cancel{\tau}\right)=-\Theta_{2}\left(k,\cancel{\tau}\right)
\end{equation}
and, finally,

\begin{equation}
 \Lambda\left(k,\cancel{\tau}\right) = \det\left(A'\right)\mathcal{I}'\left[\breve{\Psi}_{\mathrm{t}}\right]=-\Theta\left(k,\cancel{\tau}\right),
\end{equation}
as required.
\end{proof}

\section{The $\mathcal{G}$ function}\label{app:gam}

For $\Gamma\left(k\right)$ and $\mathcal{G}\left(k\right)$ as defined in (\ref{eq:COMP:gam}) and (\ref{eq:FORM:calG}) respectively, we claim that (\ref{eq:FORM:GAMAGO}) holds.

\begin{proof}
To prove (\ref{eq:FORM:GAMAGO}), inspection of (\ref{eq:FORM:calG}) shows that it is sufficient to prove

\begin{equation}\label{eq:APG:apr}
 \mathcal{H}=0,
\end{equation}
where we have defined

\begin{equation}\label{eq:APG:td}
\mathcal{H}=\left[\tilde{\mathcal{A}}\left(k\right)+\mathcal{B}\left(k\right)\right]\tilde{\mathcal{A}}\left(k\right)+\mathcal{A}\left(k\right)\mathcal{C}\left(k\right)-\left[2\tilde{\mathcal{A}}\left(k\right)+\mathcal{B}\left(k\right)\right]\Gamma\left(k\right).
\end{equation}
Henceforth, it will be convenient to consider only the case $M\geq3$ in (\ref{eq:INT:trialwaveEXPL}). It can be shown that the result (\ref{eq:APG:apr}) is satisfied for $M<3$ by explicitly evaluating expressions for $\tilde{\mathcal{A}}\left(k\right)$, $\mathcal{A}\left(k\right)$, $\mathcal{B}\left(k\right)$, $\mathcal{C}\left(k\right)$ and $\Gamma\left(k\right)$. Throughout, we will implicitly make use of the Hermiticity properties (\ref{eq:INT:HERM2})-(\ref{eq:INT:HERM5}).

First, using (\ref{eq:COMP:detzero}), (\ref{eq:COMP:detpi}), (\ref{eq:COMP:dettpi}) and (\ref{eq:COMP:detpi4}), we obtain
 
\begin{eqnarray}
\label{eq:APG:AC}\fl\mathcal{A}\left(k\right)\mathcal{C}\left(k\right)&=&\mathcal{P}^{2}\left(k\right)\langle S\psi_{\mathrm{G}},S\psi_{\mathrm{G}}\rangle\langle C\psi_{\mathrm{G}},C\psi_{\mathrm{G}}\rangle\\\nonumber\fl
&+&\mathcal{P}\left(k\right)\langle S\psi_{\mathrm{G}},S\psi_{\mathrm{G}}\rangle\sum_{i=1}^{M}\sum_{j=1}^{M}\mathcal{S}_{ij}\left(k\right)\langle C\psi_{\mathrm{G}},\chi_{i}\rangle\langle C\psi_{\mathrm{G}},\chi_{j}\rangle\\\nonumber\fl
&+&\mathcal{P}\left(k\right)\langle C\psi_{\mathrm{G}},C\psi_{\mathrm{G}}\rangle\sum_{i=1}^{M}\sum_{j=1}^{M}\mathcal{S}_{ij}\left(k\right)\langle S\psi_{\mathrm{G}},\chi_{i}\rangle\langle S\psi_{\mathrm{G}},\chi_{j}\rangle\\\nonumber\fl
&+&\sum_{i=1}^{M}\sum_{j=1}^{M}\sum_{p=1}^{M}\sum_{q=1}^{M}\mathcal{S}_{ij}\left(k\right)\mathcal{S}_{pq}\left(k\right)\langle S\psi_{\mathrm{G}},\chi_{i}\rangle\langle S\psi_{\mathrm{G}},\chi_{j}\rangle\langle C\psi_{\mathrm{G}},\chi_{p}\rangle\langle C\psi_{\mathrm{G}},\chi_{q}\rangle
\end{eqnarray} 
and
 
\begin{equation}\label{eq:APG:AB}
\fl\tilde{\mathcal{A}}\left(k\right)+\mathcal{B}\left(k\right)=-\mathcal{P}\left(k\right)\langle C\psi_{\mathrm{G}},S\psi_{\mathrm{G}}\rangle-\sum_{i=1}^{M}\sum_{j=1}^{M}\mathcal{S}_{ij}\left(k\right)\langle C\psi_{\mathrm{G}},\chi_{i}\rangle\langle S\psi_{\mathrm{G}},\chi_{j}\rangle.
\end{equation}
Next, in the nomenclature of \ref{app:theta}, we note that

\begin{equation}\label{eq:APG:rel1}
 \mathcal{P}\left(k\right)=\det\left(A^{\left(1\right)}_{\left(1\right)}\right)=\det\left(X\right)
\end{equation}
and

\begin{equation}\label{eq:APG:rel2}
 \mathcal{S}_{ij}\left(k\right)=\mathcal{S}_{ji}\left(k\right)=\left(-1\right)^{i+j+1}\det\left(X^{\left(i\right)}_{\left(j\right)}\right)=\left(-1\right)^{i+j+1}\det\left(X^{\left(j\right)}_{\left(i\right)}\right).
\end{equation}
Combining (\ref{eq:COMP:dettpi}), (\ref{eq:APG:AC}) and (\ref{eq:APG:AB}), as well as using (\ref{eq:APT:A5}), (\ref{eq:APT:A13}), (\ref{eq:APG:rel1}) and (\ref{eq:APG:rel2}), after a little work we can write

\begin{eqnarray}\label{eq:APG:lhs} \fl\left[\tilde{\mathcal{A}}\left(k\right)+\mathcal{B}\left(k\right)\right]\tilde{\mathcal{A}}\left(k\right)+\mathcal{A}\left(k\right)\mathcal{C}\left(k\right)=\mathcal{P}\left(k\right)\left[\Theta_{0}\left(k\right)+\Theta_{1}\left(k\right)\right]\\\nonumber
+\sum_{i=1}^{M}\sum_{j=1}^{M}\sum_{p=1}^{M}\sum_{q=1}^{M}\mathcal{S}_{ij}\left(k\right)\mathcal{S}_{pq}\left(k\right)\langle S\psi_{\mathrm{G}},\chi_{i}\rangle\langle S\psi_{\mathrm{G}},\chi_{j}\rangle\langle C\psi_{\mathrm{G}},\chi_{p}\rangle\langle C\psi_{\mathrm{G}},\chi_{q}\rangle\\\nonumber
-\sum_{i=1}^{M}\sum_{j=1}^{M}\sum_{p=1}^{M}\sum_{q=1}^{M}\mathcal{S}_{ij}\left(k\right)\mathcal{S}_{pq}\left(k\right)\langle S\psi_{\mathrm{G}},\chi_{i}\rangle\langle C\psi_{\mathrm{G}},\chi_{j}\rangle\langle C\psi_{\mathrm{G}},\chi_{p}\rangle\langle S\psi_{\mathrm{G}},\chi_{q}\rangle.
\end{eqnarray}
Next, summing (\ref{eq:COMP:dettpi}) and (\ref{eq:APG:AB}), we find

\begin{equation}
2\tilde{\mathcal{A}}\left(k\right)+\mathcal{B}\left(k\right)=\mathcal{P}\left(k\right)\tilde{k},
\end{equation}
where we have used (\ref{eq:INT:sccs}), together with the fact that $\mathcal{S}_{ij}\left(k\right)=\mathcal{S}_{ji}\left(k\right)$. Recalling (\ref{eq:COMP:gam}), it is clear that 

\begin{equation}\label{eq:APG:GT}
 \left[2\tilde{\mathcal{A}}\left(k\right)+\mathcal{B}\left(k\right)\right]\Gamma\left(k\right)=\mathcal{P}\left(k\right)\Theta\left(k\right).
\end{equation}
Using (\ref{eq:APT:A14}), (\ref{eq:APT:finalresult}), (\ref{eq:APG:lhs}) and (\ref{eq:APG:GT}), after some cancellation we can then rewrite (\ref{eq:APG:td}) as 

\begin{equation}
 \fl\mathcal{H}=\sum_{i=1}^M\sum_{j=1}^M\sum_{p=1}^M\sum_{q=1}^M \mathcal{T} \left(-1\right)^{i+j+p+q} \langle S\psi_{\mathrm{G}},\chi_{i}\rangle \langle S\psi_{\mathrm{G}},\chi_{j}\rangle \langle C\psi_{\mathrm{G}},\chi_{p}\rangle \langle C\psi_{\mathrm{G}},\chi_{q}\rangle,
\end{equation}
where, using (\ref{eq:APG:rel1}) and (\ref{eq:APG:rel2}), we have defined 

\begin{eqnarray}\nonumber
\mathcal{T}&=&\det\left(X^{\left(i\right)}_{\left(j\right)}\right)\det\left(X^{\left(p\right)}_{\left(q\right)}\right)-\det\left(X^{\left(p\right)}_{\left(j\right)}\right)\det\left(X^{\left(i\right)}_{\left(q\right)}\right)\\\label{eq:APG:Z}
&-&\left(-1\right)^{\sigma_{ip}+\sigma_{jq}}\det\left(X\right)\det\left(X^{\left(i,p\right)}_{\left(j,q\right)}\right).
\end{eqnarray}
If $i=p$ or $j=q$, it follows trivially from the definition of $X^{\left(i,p\right)}_{\left(j,q\right)}$ given in \ref{app:theta} that $\mathcal{T}=0$. When $i<p$ and $j<q$, we make use of the following result,

\begin{equation}\label{eq:APG:LewisCarrollIdentity}\fl
 \det\left(X\right)\det\left(X^{\left(i,p\right)}_{\left(j,q\right)}\right)=\det\left(X^{\left(i\right)}_{\left(j\right)}\right)\det\left(X^{\left(p\right)}_{\left(q\right)}\right)-\det\left(X^{\left(p\right)}_{\left(j\right)}\right)\det\left(X^{\left(i\right)}_{\left(q\right)}\right).
\end{equation}
Commonly known as the Lewis Carroll identity after its role in Dodgson condensation \cite{Dodgson1860}, Fomin and Zelevinsky \cite{Fomin2000} point out that (\ref{eq:APG:LewisCarrollIdentity}) was, in fact, proved earlier by Desnanot (see, for example, \cite{Muir1906}). It is easily seen to generalize to any $i\neq p$ and $j\neq q$ by multiplying the left hand side of (\ref{eq:APG:LewisCarrollIdentity}) by a factor of $\left(-1\right)^{\sigma_{ip}+\sigma_{jq}}$. Hence, from inspection of (\ref{eq:APG:Z}) we see that $\mathcal{T}$ is identically zero and, since

\begin{equation}
 \mathcal{T}=0 \Rightarrow \mathcal{H}=0 \Rightarrow \mathcal{G}\left(k\right)=\Gamma^2\left(k\right),
\end{equation}
the required result (\ref{eq:FORM:GAMAGO}) follows.
\end{proof}

\section{Derivations of (\ref{eq:COMP:retand}) and (\ref{eq:COMP:imtand})}\label{app:retan}

We consider a calculation under the same conditions on $k$ as outlined at the beginning of subsection \ref{s:equiv}. Under these circumstances, we claim that (\ref{eq:COMP:retand}) and (\ref{eq:COMP:imtand}) hold.

To derive (\ref{eq:COMP:retand}), we begin by defining the following functions,

\begin{eqnarray}\label{eq:retand:p1}
\mathcal{D}\left(k\right)&=&2\tilde{\mathcal{A}}\left(k\right)+\mathcal{B}\left(k\right)-2\Gamma\left(k\right),\\
\mathcal{E}\left(k\right)&=&\mathcal{A}\left(k\right)+\mathcal{C}\left(k\right),\\
p\left(\tau;k\right)&=&\mathcal{D}\left(k\right)v\left(\tau;k\right)-\mathcal{E}\left(k\right)u\left(\tau;k\right),\\
q\left(\tau;k\right)&=&\mathcal{D}\left(k\right)u\left(\tau;k\right)+\mathcal{E}\left(k\right)v\left(\tau;k\right),
\end{eqnarray}
where $u\left(\tau;k\right)$ and $v\left(\tau;k\right)$ are as in (\ref{eq:COMP:u2}) and (\ref{eq:COMP:v2}), respectively. Using (\ref{eq:FORM:tanpreal}), (\ref{eq:COMP:fintan}) and (\ref{eq:COMP:tandiff}), we then find

\begin{equation}\label{eq:retand:p2}\fl
\Re\left[\tan\left(\eta_{\mathrm{v}}^{\left(i\right)}-\eta'_{\mathrm{v}}-\tau_{i}+\tau\right)\right]=\frac{\Re\left[p\left(\tau;k\right)\right]\Re\left[q\left(\tau;k\right)\right]+\Im\left[p\left(\tau;k\right)\right]\Im\left[q\left(\tau;k\right)\right]
}{\Re\left[q\left(\tau;k\right)\right]\Re\left[q\left(\tau;k\right)\right]+\Im\left[q\left(\tau;k\right)\right]\Im\left[q\left(\tau;k\right)\right]}.
\end{equation}
We further define

\begin{eqnarray}
d_{1}\left(\tau;k\right)&=&\left[\mathcal{A}\left(k\right)-\mathcal{C}\left(k\right)\right]\cos\left(2\tau\right)-\mathcal{B}\left(k\right)\sin\left(2\tau\right),\\
d_{2}\left(\tau;k\right)&=&\left[\mathcal{A}\left(k\right)-\mathcal{C}\left(k\right)\right]\sin\left(2\tau\right)+\mathcal{B}\left(k\right)\cos\left(2\tau\right).
\end{eqnarray}
After a little work, we then find

\begin{eqnarray}\nonumber\fl&&
\Re\left[p\left(\tau;k\right)\right]\Re\left[q\left(\tau;k\right)\right]+\Im\left[p\left(\tau;k\right)\right]\Im\left[q\left(\tau;k\right)\right]\\\nonumber\fl
&=&\left[\mathcal{D}^{2}\left(k\right)-\mathcal{E}^{2}\left(k\right)\right]\left[\Re\left[u\left(\tau;k\right)\right]d_{1}\left(\tau;k\right)-\Im\left[u\left(\tau;k\right)\right]d_{2}\left(\tau;k\right)\right]\\\label{eq:retand:b1}\fl
&+&\mathcal{D}\left(k\right)\mathcal{E}\left(k\right)\left[d_{1}^{2}\left(\tau;k\right)+d_{2}^{2}\left(\tau;k\right)+2\Im\left[u\left(\tau;k\right)\right]d_{1}\left(\tau;k\right)+2\Re\left[u\left(\tau;k\right)\right]d_{2}\left(\tau;k\right)\right]
\end{eqnarray}
and

\begin{eqnarray}
\nonumber\fl&&
\Re\left[q\left(\tau;k\right)\right]\Re\left[q\left(\tau;k\right)\right]+\Im\left[q\left(\tau;k\right)\right]\Im\left[q\left(\tau;k\right)\right]\\\nonumber\fl
&=&2\mathcal{D}\left(k\right)\mathcal{E}\left(k\right)\left[\Re\left[u\left(\tau;k\right)\right]d_{1}\left(\tau;k\right)-\Im\left[u\left(\tau;k\right)\right]d_{2}\left(\tau;k\right)\right]\\\nonumber\fl
&+&\mathcal{E}^{2}\left(k\right)\left[d_{1}^{2}\left(\tau;k\right)+d_{2}^{2}\left(\tau;k\right)+2\Im\left[u\left(\tau;k\right)\right]d_{1}\left(\tau;k\right)+2\Re\left[u\left(\tau;k\right)\right]d_{2}\left(\tau;k\right)\right]\\\fl
&+&\left[\mathcal{D}^{2}\left(k\right)+\mathcal{E}^{2}\left(k\right)\right]\left[\Re\left[u\left(\tau;k\right)\right]\Re\left[u\left(\tau;k\right)\right]+\Im\left[u\left(\tau;k\right)\right]\Im\left[u\left(\tau;k\right)\right]\right],
\end{eqnarray}
where we have made use of (\ref{eq:COMP:v2}) to eliminate $v\left(\tau;k\right)$. Next, we define

\begin{eqnarray}
\fl\mathcal{M}\left(k\right)&=&\left[\tilde{\mathcal{A}}\left(k\right)-\Gamma\left(k\right)\right]\left[\mathcal{C}\left(k\right)-\mathcal{A}\left(k\right)\right]+\mathcal{B}\left(k\right)\mathcal{C}\left(k\right),\\
\fl\mathcal{N}\left(k\right)&=&\mathcal{B}\left(k\right)\left[\tilde{\mathcal{A}}\left(k\right)-\Gamma\left(k\right)\right]+\frac{\mathcal{A}^{2}\left(k\right)+\mathcal{B}^{2}\left(k\right)-\mathcal{C}^{2}\left(k\right)}{2},
\end{eqnarray}
so that, after some laborious but elementary operations, using (\ref{eq:COMP:u2}) we obtain

\begin{equation}\fl
\Re\left[u\left(\tau;k\right)\right]d_{1}\left(\tau;k\right)-\Im\left[u\left(\tau;k\right)\right]d_{2}\left(\tau;k\right)=\mathcal{M}\left(k\right)\cos\left(2\tau\right)+\mathcal{N}\left(k\right)\sin\left(2\tau\right),
\end{equation}
together with

\begin{eqnarray}\nonumber\fl
&&d_{1}^{2}\left(\tau;k\right)+d_{2}^{2}\left(\tau;k\right)+2\Im\left[u\left(\tau;k\right)\right]d_{1}\left(\tau;k\right)+2\Re\left[u\left(\tau;k\right)\right]d_{2}\left(\tau;k\right)\\\fl
&=&2\mathcal{M}\left(k\right)\sin\left(2\tau\right)-2\mathcal{N}\left(k\right)\cos\left(2\tau\right)
\end{eqnarray}
and

\begin{eqnarray}\nonumber\fl
\Re\left[u\left(\tau;k\right)\right]\Re\left[u\left(\tau;k\right)\right]+\Im\left[u\left(\tau;k\right)\right]\Im\left[u\left(\tau;k\right)\right]&=&2\mathcal{N}\left(k\right)\cos^{2}\left(\tau\right)-\mathcal{M}\left(k\right)\sin\left(2\tau\right)\\\fl
&+&\left[\tilde{\mathcal{A}}\left(k\right)-\Gamma\left(k\right)\right]^{2}+\mathcal{C}^{2}\left(k\right).
\end{eqnarray}
Next, we define

\begin{equation}\label{eq:retand:b2}
\mathcal{Z}\left(k\right)=-4\Gamma^2\left(k\right)-\mathcal{B}^{2}\left(k\right)-\left[\mathcal{A}\left(k\right)-\mathcal{C}\left(k\right)\right]^2,
\end{equation}
so that, using (\ref{eq:FORM:gsim}), (\ref{eq:FORM:x}), (\ref{eq:FORM:y}) and (\ref{eq:retand:b1})-(\ref{eq:retand:b2}), after a lengthy effort we find

\begin{eqnarray}\nonumber\fl
&&\Re\left[p\left(\tau;k\right)\right]\Re\left[q\left(\tau;k\right)\right]+\Im\left[p\left(\tau;k\right)\right]\Im\left[q\left(\tau;k\right)\right]\\\fl
&=&\mathcal{Z}\left(k\right)\left[\mathcal{X}\left(k\right)\sin\left(2\tau\right)+\mathcal{Y}\left(k\right)\cos\left(2\tau\right)\right]
\end{eqnarray}
and

\begin{eqnarray}\nonumber\fl
&&\Re\left[q\left(\tau;k\right)\right]\Re\left[q\left(\tau;k\right)\right]+\Im\left[q\left(\tau;k\right)\right]\Im\left[q\left(\tau;k\right)\right]\\\label{eq:COMP:finr1}\fl
&=&\mathcal{Z}\left(k\right)\left[2\mathcal{X}\left(k\right)\cos^2\left(\tau\right)-\mathcal{Y}\left(k\right)\sin\left(2\tau\right)-\left[\tilde{\mathcal{A}}\left(k\right)-\Gamma\left(k\right)\right]^2-\mathcal{A}^2\left(k\right)\right].
\end{eqnarray}
Finally, after some work, examination of (\ref{eq:COMP:f}) and (\ref{eq:COMP:g}) yields, 

\begin{equation}\label{eq:COMP:finr2}\fl
-f^{2}\left(\tau;k\right)-g^2\left(\tau;k\right)=2\mathcal{X}\left(k\right)\cos^2\left(\tau\right)-\mathcal{Y}\left(k\right)\sin\left(2\tau\right)-\left[\tilde{\mathcal{A}}\left(k\right)-\Gamma\left(k\right)\right]^2-\mathcal{A}^2\left(k\right).
\end{equation}
The required result (\ref{eq:COMP:retand}) then follows immediately from (\ref{eq:COMP:retantop}), (\ref{eq:COMP:retanbot}), (\ref{eq:retand:p2}), (\ref{eq:COMP:finr1}), (\ref{eq:COMP:finr2}) and cancellation of $\mathcal{Z}\left(k\right)$. Since we have assumed $k\neq k_{\mathrm{g}}$ and $k\neq k_{\mathrm{h}}$, it follows directly from (\ref{eq:FORM:x}), (\ref{eq:FORM:y}) and (\ref{eq:retand:b2}) that $\mathcal{Z}\left(k\right)<0$.

The result (\ref{eq:COMP:imtand}) is derived by first noting that

\begin{equation}\label{eq:retand:p3}\fl
\Im\left[\tan\left(\eta_{\mathrm{v}}^{\left(i\right)}-\eta'_{\mathrm{v}}-\tau_{i}+\tau\right)\right]=\frac{\Im\left[p\left(\tau;k\right)\right]\Re\left[q\left(\tau;k\right)\right]-\Im\left[q\left(\tau;k\right)\right]\Re\left[p\left(\tau;k\right)\right]
}{\Re\left[q\left(\tau;k\right)\right]\Re\left[q\left(\tau;k\right)\right]+\Im\left[q\left(\tau;k\right)\right]\Im\left[q\left(\tau;k\right)\right]}.
\end{equation}
By a method analogous to that used to derive (\ref{eq:COMP:retand}), we eventually find that

\begin{equation}
\Im\left[p\left(\tau;k\right)\right]\Re\left[q\left(\tau;k\right)\right]-\Im\left[q\left(\tau;k\right)\right]\Re\left[p\left(\tau;k\right)\right]=\mathcal{Z}\left(k\right)\Gamma^2\left(k\right).
\end{equation}
The required result (\ref{eq:COMP:imtand}) then follows in the obvious way.

\end{document}